\documentclass[10pt]{article}

\usepackage[pdftex]{graphicx}
\graphicspath{{./figures/}}
\usepackage{csquotes}
\usepackage{xcolor}
\usepackage{subcaption}
\usepackage{graphicx}
\usepackage{float}
\usepackage[T1]{fontenc}
\usepackage{hyperref}
\usepackage{csagh}
\begin{document}

\begin{opening}

\title{Reconstruction of muon bundles in KM3NeT detectors using machine learning methods}

\author[{
AstroCeNT, Nicolaus Copernicus Astronomical Center, Polish Academy of Sciences, Rektorska 4, 00-614 Warsaw, Poland, pkalaczynski@camk.edu.pl \newline
Center of Excellence in Artificial Intelligence, AGH University of Krakow, Al. Mickiewicza 30, 30-059 Krakow, Poland, pkalaczynski@agh.edu.pl
}]{Piotr Kalaczy{\'n}ski}

\begin{abstract}
The KM3NeT Collaboration is installing the ARCA and ORCA neutrino detectors at the bottom of the Mediterranean Sea. The focus of ARCA is neutrino astronomy, while ORCA is optimised for neutrino oscillation studies. Both detectors are already operational in their intermediate states and collect valuable data, including the measurements of the muons produced by cosmic ray interactions in the atmosphere. This work explores the potential of machine learning models for the reconstruction of muon bundles, which are multi-muon events. For this, data collected with intermediate detector configurations of ARCA and ORCA was used in addition to simulated data from the envisaged final configurations of those detectors. Prediction of the total number of muons in a bundle as well as their total energy and even the energy of the primary cosmic ray is presented.

\end{abstract}

\keywords{machine learning, cosmic ray, KM3NeT, MUPAGE, CORSIKA, muon, multiplicity, neutrino, primary, energy, LightGBM, GZK, PMT, Cherenkov}

\end{opening}

\section{Introduction}
\label{Intro}

Atmospheric muons produced in cosmic ray (see Sec. \ref{CR}) interactions to many may seem a \textit{terra cognita}. Upon closer examination it becomes clear there is still much to be explored. The most prominent example is the (in-)famous Muon Puzzle \cite{Muon_Puzzle,KM3NeT_Muons_Andrey}: experiments measuring the muon flux consistently report a deficit in muon events observed at high energies and high multiplicities (number of muons from a single cosmic ray interaction). Uncertainties alone cannot explain the discrepancy between the measurements and current state-of-the-art theoretical predictions. The KM3NeT observatory contributes to a better understanding of the problem by measuring the muon flux at the bottom of the Mediterranean Sea. A simulation workflow has been established to study the properties of muon bundles --- groups of muons originating from the same primary cosmic ray. With a dedicated muon simulation, it was possible to train a machine learning model to reconstruct several important observables, such as the total number of muons in a bundle, their total energy and the energy of the primary cosmic ray. Measuring those quantities can help to pinpoint the shortcomings of the current theoretical models used to compute the muon flux.

\section{Cosmic rays}
\label{CR}

The existence of cosmic radiation was discovered by Victor Hess in 1912 following a series of balloon-borne radiation measurements \cite{Hess:1912srp}. Since then, a wide range of experiments have embarked on a quest for better understanding of the nature and origin of cosmic rays (CRs) and pushed to improve the precision of CR measurements \cite{AMS-02,AUGER,KASKADE,PAMELA,TA}.

\begin{figure}[H]
    \centering
    \includegraphics[width=1.0\linewidth]{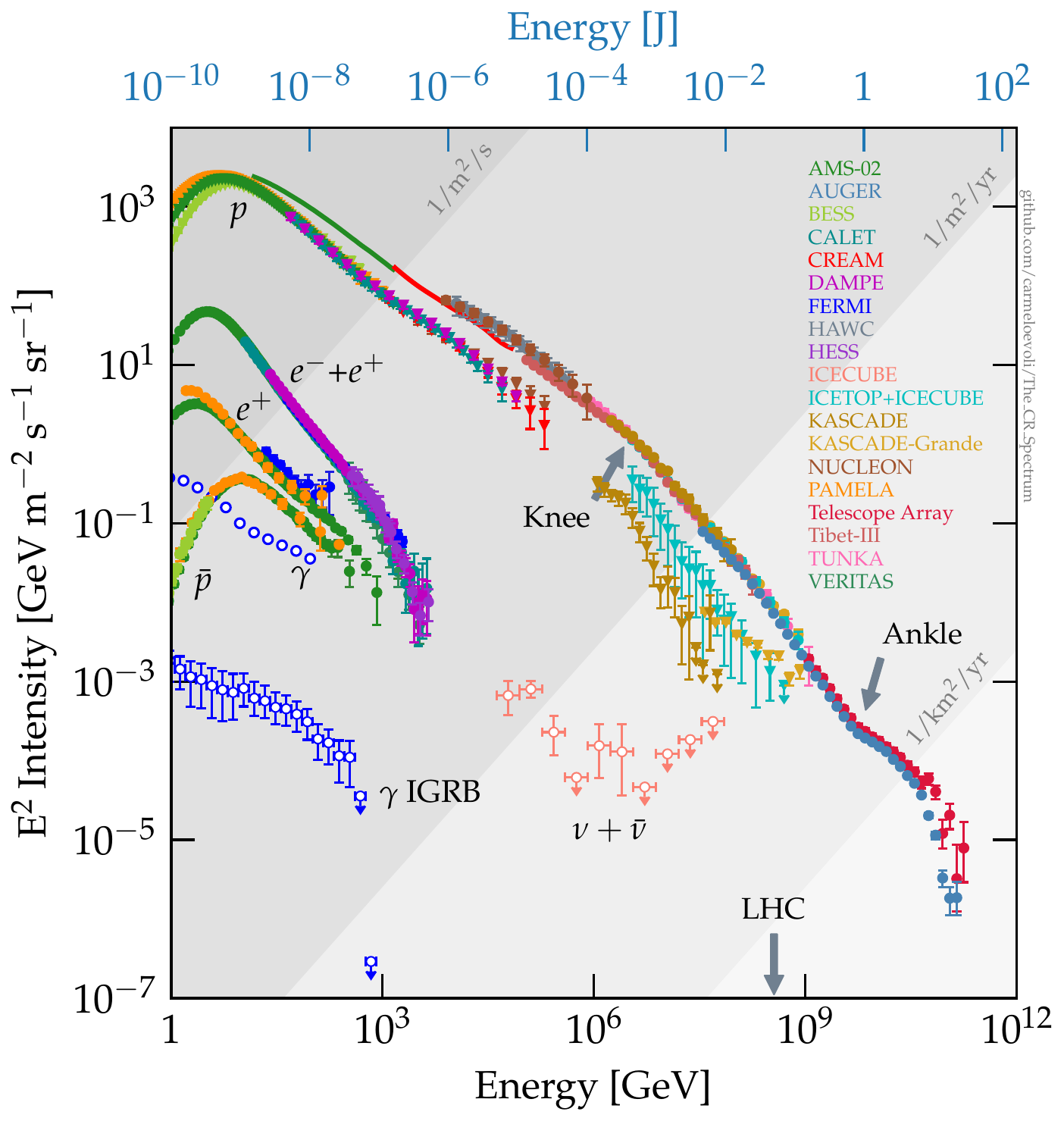}
    \caption{Energy spectra of various CR primary particles measured by several experiments. The "LHC" arrow indicates the kinetic energy achievable with the Large Hadron Collider (as of 2023). The figure was adapted from \cite{The_CR_spectrum}.}
    \label{fig:cr_spectrum}
\end{figure}

A CR is a high-energy particle or atomic nucleus arriving at Earth from outer space. As shown in Fig. \ref{fig:cr_spectrum}, the vast majority of primary CRs are protons, followed by electrons, positrons and photons. A tiny contribution from the cosmic neutrinos has been observed as well. Direct measurements of primary CRs are only possible with detectors placed at extremely high altitudes. What can be observed at the Earth's surface and below are the products of CR interactions with the atmospheric molecules. Such collisions result in a cascade of particles called an extensive air shower (EAS). A typical EAS starts at an altitude of approximately 30\,km; some of the secondary particles will reach sea level and even some distance underground or underwater. The most penetrant secondaries are muons and neutrinos, which are studied using the KM3NeT telescopes.

\section{The KM3NeT experiment}
\label{KM3NeT}

\begin{figure}[H]
    \centering
    \includegraphics[width=1.0\linewidth]{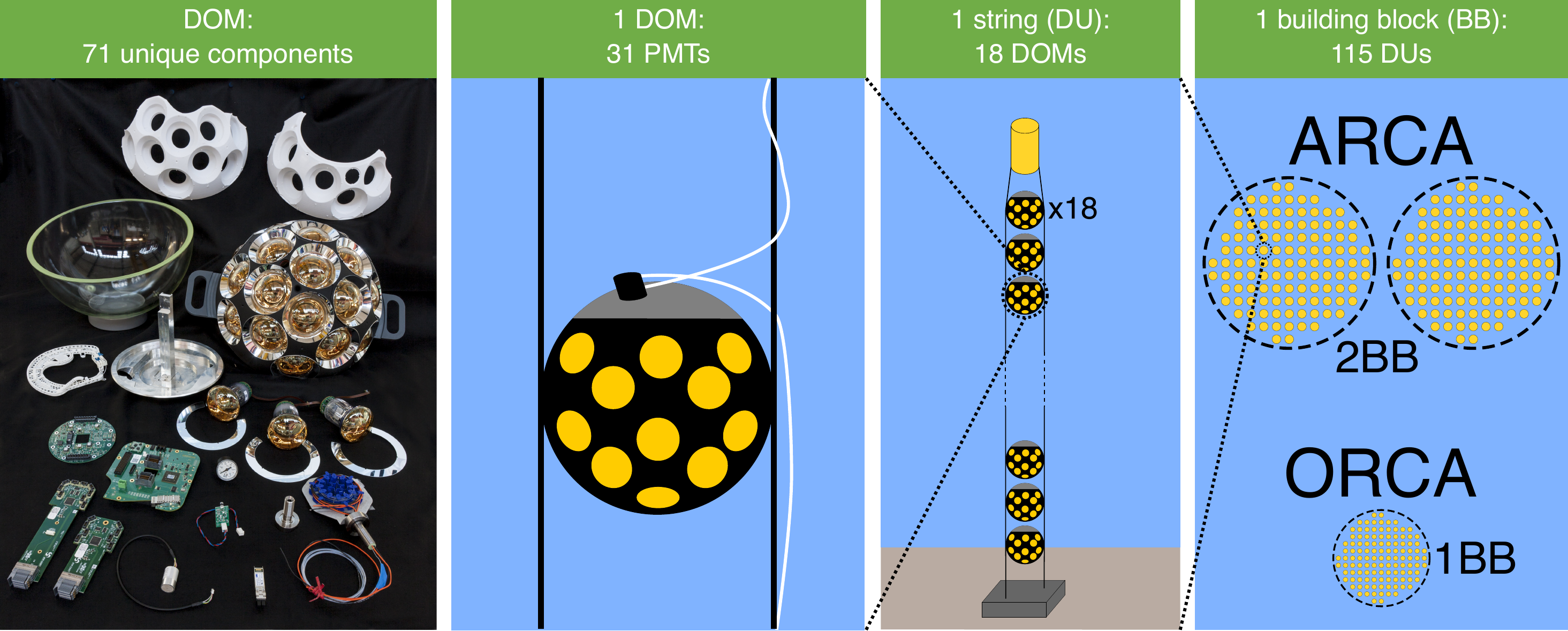}
    \caption{Pictorial summary of the design of KM3NeT neutrino telescopes. From left to right: the components of an optical module, the module integrated into a detection unit, an artist's impression of a detection unit and a building block.}
    \label{fig:km3net_desig}
\end{figure}

KM3NeT \cite{KM3NeT_LoI} is a network of two neutrino telescopes at the bottom of the Mediterranean Sea, currently under construction: ARCA near Portopalo di Capo Passero in Italy and ORCA offshore Toulon in France. They share the same hardware, however, their physics focus and thus geometry differ. ARCA is located at a depth of 3500\,m and is the more sparse of the two detectors, allowing for the study of cosmic neutrinos in the TeV-PeV energy range. ORCA is placed shallower, at 2500\,m underwater. Its more dense instrumentation allows for precise studies of atmospheric neutrino oscillations in the GeV-TeV energy range. The design of the KM3NeT detectors is summarised in Fig. \ref{fig:km3net_desig}. The optical modules are composed of many elements; the most important are the photomultiplier tubes (PMTs), which measure the Cherenkov light. Muons and other electrically charged products of neutrino interactions travelling faster than light in water cause emission of Cherenkov photons by the water molecules upon deexcitation. Each optical module contains 31 PMTs facing different directions, which allows for directional reconstruction \cite{Multi-PMT-KM3NeT-DOM}. A vertical line composed of 18 buoyant DOMs, anchored to the seabed and kept straight by a buoy is called a detection unit. A building block of a detector consists of 115 detection units. ORCA will comprise a single building block, while for ARCA two building blocks are foreseen. Separation into two separate blocks instead of a single larger one has been driven by cost-efficiency considerations rather than a specific physics goal.

The main interest of KM3NeT is to observe neutrinos. However, most of the collected data consists of muons produced in EASs. The muons do not only constitute the main background for neutrino measurements due to a similar signal produced in the detector but also present a valuable opportunity for indirect study of the primary CRs from which they originate.

\section{Muon simulation}
\label{sim}

In KM3NeT, two standard muon simulation tools are developed: MUPAGE \cite{MUPAGE} and CORSIKA \cite{CORSIKA}. Both were used for this work and are briefly described in what follows.

MUPAGE (atmospheric MUons from PArametric formulas: a fast GEnerator for neutrino telescopes) is a code inherited by KM3NeT from the ANTARES experiment. It extensively makes use of parametrised functions, which can be tuned to experimental data and simulations. The original tuning was performed based on the HEMAS \cite{HEMAS} Monte Carlo (MC) simulation code and data from the MACRO experiment at Gran Sasso in Italy \cite{MACRO}. The output of MUPAGE consists of muon bundle events sampled on a cylindrical volume surrounding the detector, called the can.

CORSIKA (COsmic Ray SImulations for KAscade) originated as an MC simulation code for the KASKADE experiment \cite{KASKADE}. With time, it evolved into a state-of-the-art solution for simulating cosmic ray air showers. The code generates EASs starting from the very beginning, i.e. from the first interaction of the primary particle in the atmosphere. It includes a plethora of customization possibilities such as the geometry of the problem, the assumed atmospheric density profile, the choice of low- and high-energy hadronic interaction models and more. Although currently only used for muon simulations by KM3NeT, CORSIKA is not limited to this. All particles produced in the shower are available to the user at a specified observation level (sea level in the case of KM3NeT simulations).

\begin{figure}[H]
    \centering
    \includegraphics[width=1.0\linewidth]{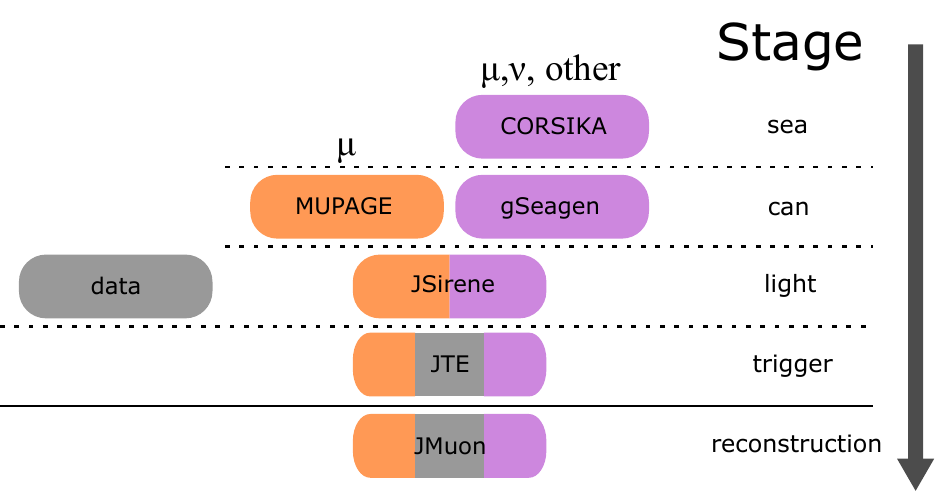}
    \caption{
    The processing chain for muon simulations in KM3NeT. The colour coding distinguishes three possibilities: CORSIKA simulation marked in purple, MUPAGE simulation in orange and experimental data indicated by the grey colour. The order of execution is from top to bottom, along the vertical arrow. At the trigger stage, both simulations and experimental data can be compared against each other.
    }
    \label{fig:processing_chain}
\end{figure}

In Fig. \ref{fig:processing_chain}, the processing chain of the muon simulation is shown. As can be seen, the starting points differ, however, both simulation codes and the experimental data go through certain common stages. The sea level is unique to CORSIKA simulation and to transport the muons to the can, the \href{https://git.km3net.de/opensource/gseagen/}{gSeaGen} \cite{gSeaGen} software was adapted by the author of this work. It is the subject of an upcoming paper "gSeaGen code by KM3NeT: an efficient tool to propagate muons simulated with CORSIKA". JSirene \cite{JSirene} is an application developed to simulate photon emission and detection by the PMTs, including the signal from environmental background. The JTE code selects potentially interesting events, i.e. performs triggering. Finally, the observables are reconstructed for each event using a specialized algorithm. Here, the standard KM3NeT tool for muon track reconstruction is JMuon. It is a multi-stage fit based on a maximum likelihood approach, assuming a single muon hypothesis. This paper presents a novel reconstruction method utilizing machine learning techniques to predict the muon bundle energy as well as the energy of the primary cosmic ray and the muon multiplicity.

\section{Reconstruction of muon bundle properties}
\label{reco}
In this section, the reconstruction of three muon bundle observables is presented: bundle energy, primary CR energy and muon multiplicity. The data used to produce the results is summarized in Tab. \ref{tab:datasets}. The reconstruction was performed separately for two detector configurations for both ARCA and ORCA. In each case, it was a full building block (ARCA115, ORCA115) and an intermediate state with 6 detection units installed (ARCA6, ORCA6).

\begin{table}[]
    \centering
    \caption{Summary of the datasets used to train and evaluate the machine learning models. The simulated data was split between training, validation and test datasets in proportion 64:16:20 and was extracted from the official KM3NeT CORSIKA simulation produced by the author of this work. The experimental data comes from the measurements taken by ARCA6 and ORCA6 detectors in 2020 and 2021.}
    \begin{tabular}{|c||c|c|c|c|}
    \cline{2-5}
    \multicolumn{1}{c|}{} & {ARCA115} & {ARCA6} & {ORCA115} & {ORCA6} \\ \hline
    Training data & 14 374 415 & 3 894 293 & 15 439 946 & 6 154 895 \\ \hline
    Validation data & 3 598 133 & \; \, 974 371 & 3 857 715  & 1 539 988 \\ \hline
    Test data & 4 489 177 & 1 215 552 & 4 825 073 & 1 923 475 \\ \hline
    Experimental data & {\color{red}$\times$}  & 3 013 982  & {\color{red}$\times$} & 4 574 887  \\ \hline
    \end{tabular}

    \label{tab:datasets}
\end{table}

The data preprocessing included extracting 46 features (described in detail in \cite{MyPhD}), event weights and target (observable) values. Each feature had its mean removed and was scaled to unit variance (standard scaling). This was done for each detector configuration before the division into training, validation and test datasets. Event weight scaling was also considered but proved not to improve the training. Before reaching the final result of reconstruction, several loops of training and validation were performed to select the best machine learning (ML) regression model for the task, pick the features that improve the performance and tune the hyperparameters (internal parameters of the model). The regressor selection was performed only for bundle energy; the result was used for primary CR energy and muon multiplicity reconstruction. The outcome for each of the observables is presented in the following subsections.

% \begin{figure}[H]
%     \centering
%     \includegraphics[width=1.0\linewidth]{figures/v6.9_ARCA115_reco_correlation_matrix_annotated.pdf}
%     \caption{Impression of the correlation matrix of all 46 features and 4 investigated targets. There are two muon multiplicities --- one has no muon selection and the other one counts only certain muons (passing the criteria specified in Sec. \ref{subsec:multiplicity}). The image is admittedly dense, with so many features.}
%     \label{fig:corr_matrix}
% \end{figure}

\subsection{Muon bundle energy}

\begin{figure}[H]
    \centering
    \includegraphics[width=1.0\linewidth]{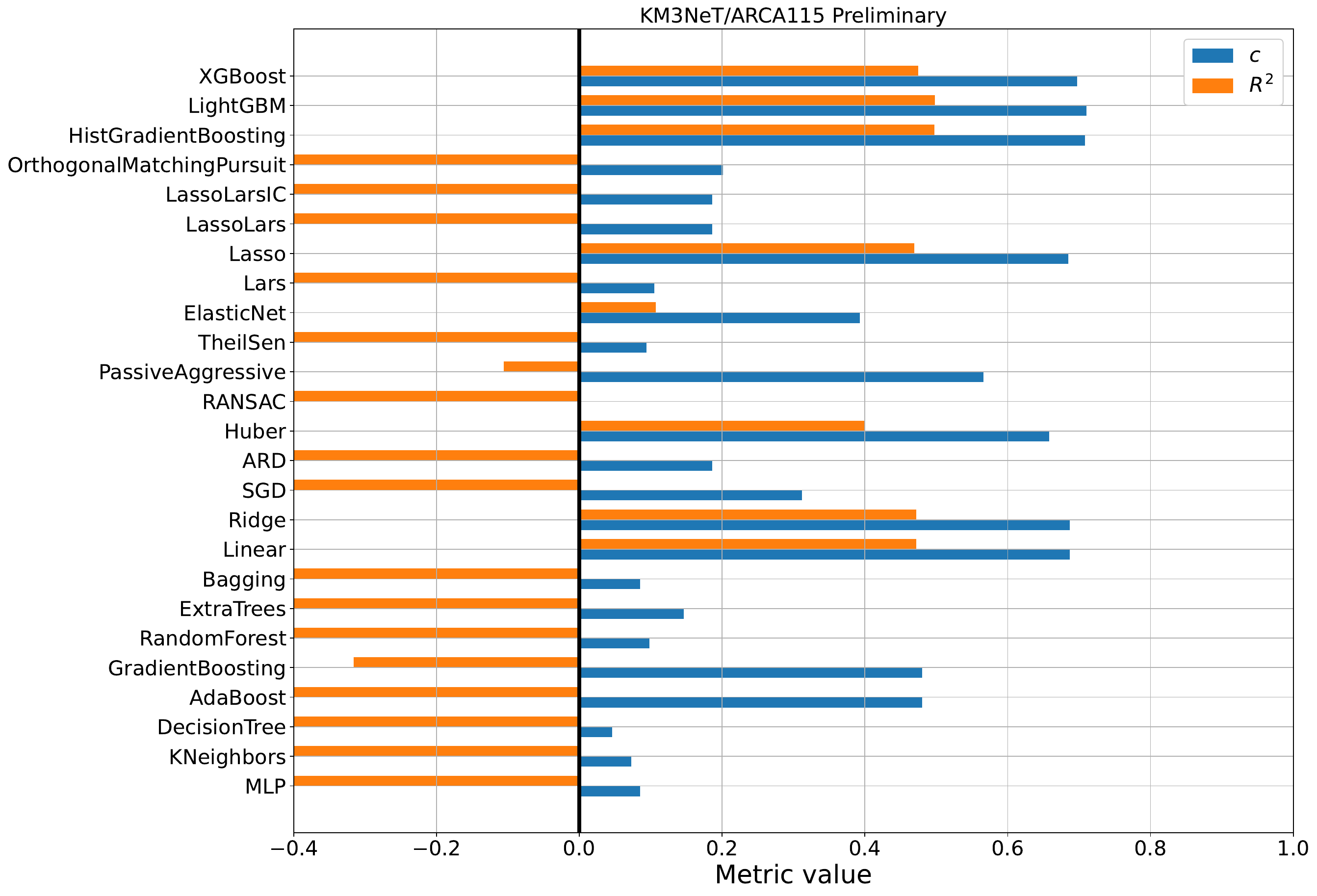}
    \caption{Different machine learning models in terms of achieved weighted $R^2$-score and weighted Pearson correlation coefficient $c$. The models were trained on a fraction (50k) of the ARCA115 training dataset because of the infeasibly long runtime and RAM memory required by some of the models.}
    \label{fig:model_selection}
\end{figure}

The first step in determining the muon bundle energy was the selection of the best ML model. A range of regression models available in the Scikit-learn library \cite{Scikit-learn}, as well as LightGBM \cite{LightGBM} and XGBoost \cite{XGBoost} were considered. The criteria for determining the best model were the weighted coefficient of determination ($R^2$-score) and weighted Pearson correlation coefficient $c$. As demonstrated in Fig. \ref{fig:model_selection}, the best performer was LightGBM, a histogram-based gradient-boosting model. Not only was it superior in terms of achieved scores, but thanks to working on the histograms rather than on the events themselves, it turned out to be one of the fastest models as well (see \cite{MyPhD}). The negative $R^2$ despite quite high $c$ value in case of some models is a result of the difference in focus of the two metrics. The $R^2$-score explicitly tests the difference between the truth and prediction, while the correlation coefficient only checks whether their relation is close to linear. Examples of 2D histograms with such negative $R^2$ can be found in the appendix of \cite{MyPhD}.

\begin{figure}[H]
    \centering
    \includegraphics[width=1.0\linewidth]{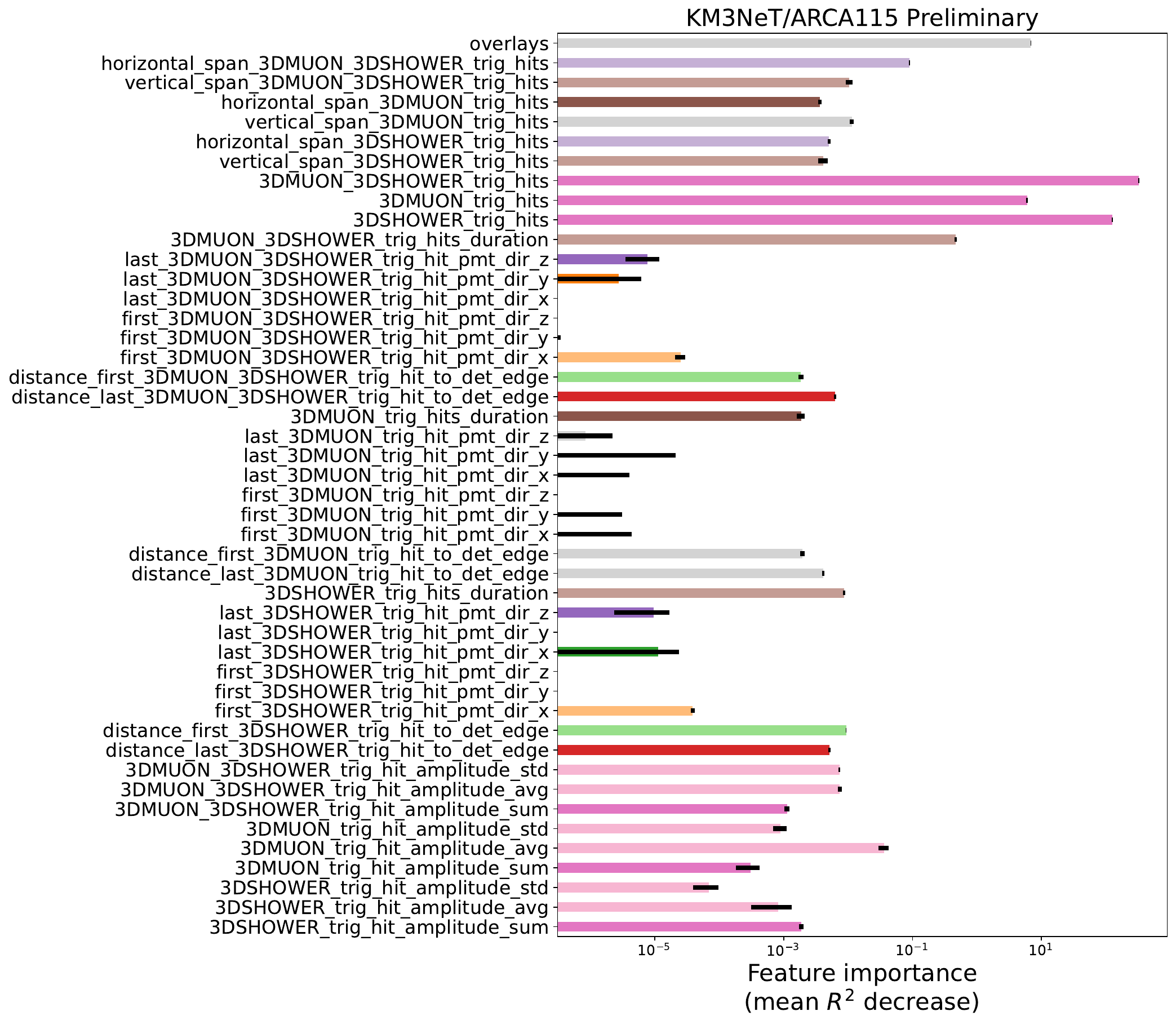}
    \caption{Feature importance computed over 10 random trials of removing a given feature. The colour coding groups features into clusters based on their inter-correlation.}
    \label{fig:feature_importance}
\end{figure}

The next step was to evaluate the usefulness of individual features. Not all were equally important and several were strongly inter-correlated, which could indicate redundancy. The feature importance was defined as the mean decrease of the $R^2$-score upon removal of a given feature. The results for ARCA115 are shown in Fig. \ref{fig:feature_importance}. Four different approaches to feature selection were considered:
\begin{enumerate}
    \item Using all available features.
    \item Taking only features with positive importance and selecting the most important feature from each cluster.
    \item Using only the single most important feature.
    \item Requiring positive importance for all features.
\end{enumerate}

\begin{figure}[H]
    \centering
    \begin{subfigure}[H]{0.49\textwidth}
         \centering
         \includegraphics[width=\textwidth]{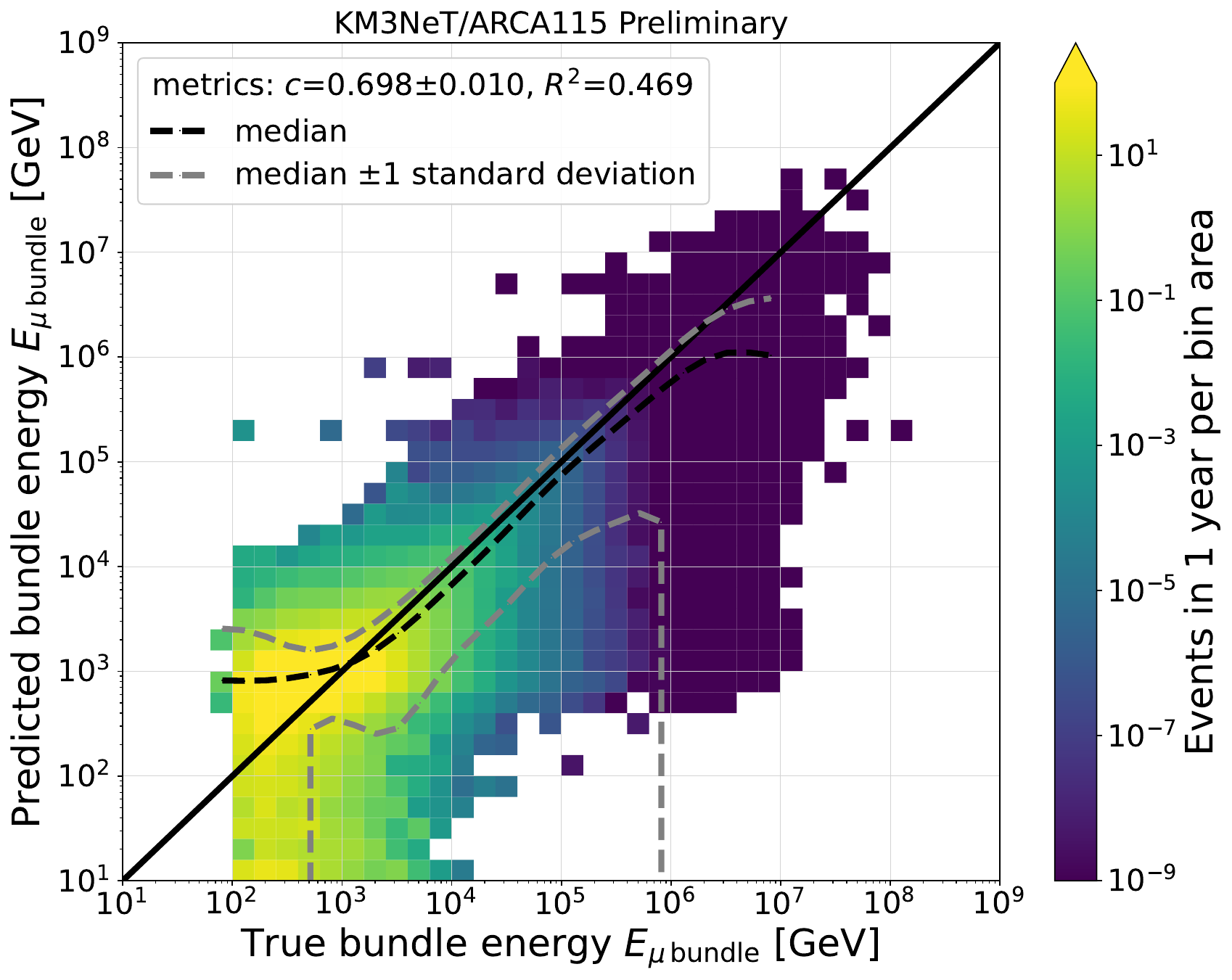}
         \caption{All features.\\ \,}
         \label{fig:all_features}
     \end{subfigure}
     \hfill
     \begin{subfigure}[H]{0.49\textwidth}
         \centering
         \includegraphics[width=\textwidth]{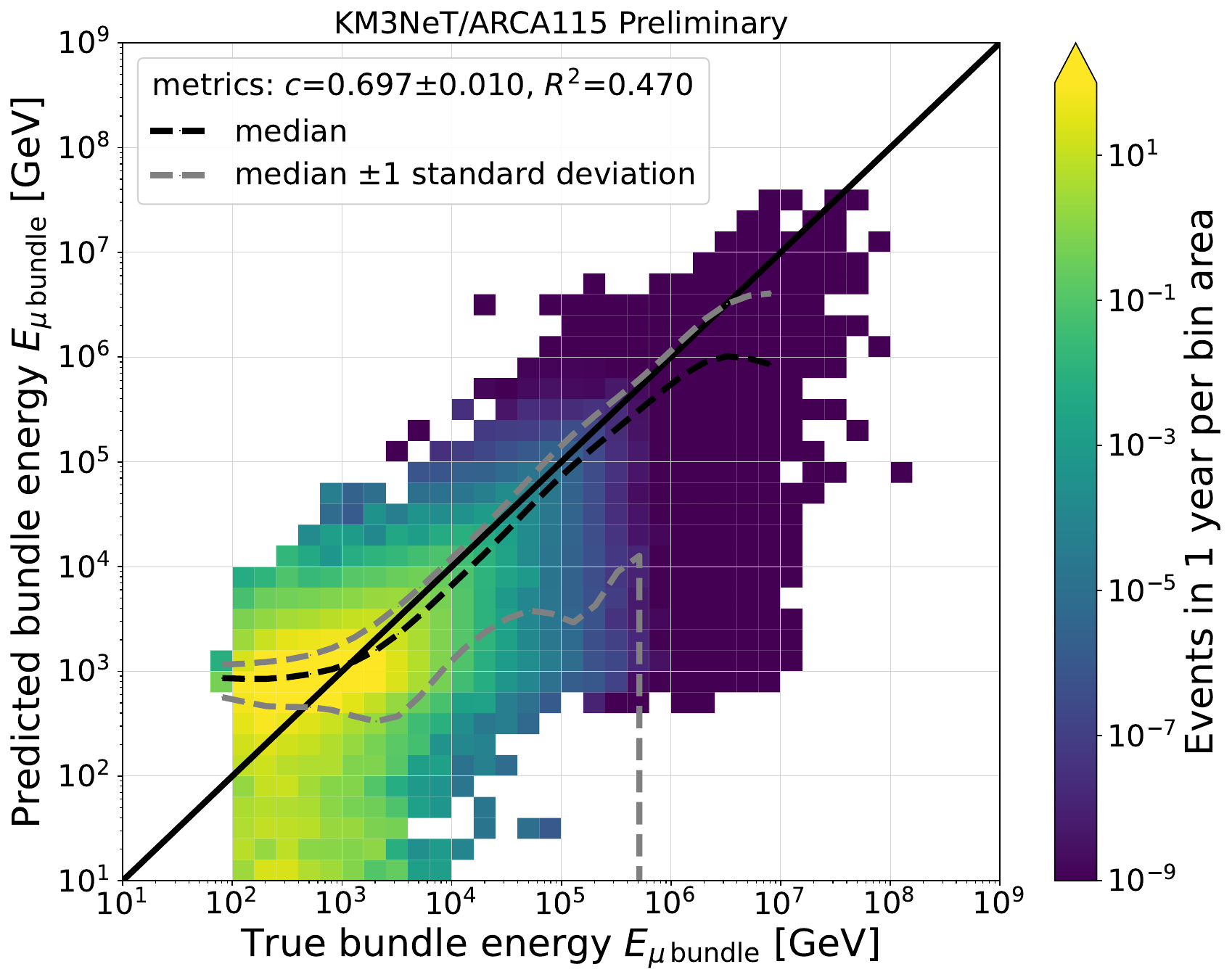}
         \caption{Selected features \\ (importance \& clustering).}
         \label{fig:selected_features_with_clustering}
     \end{subfigure}
     \newline
     \begin{subfigure}[H]{0.49\textwidth}
         \centering
         \includegraphics[width=\textwidth]{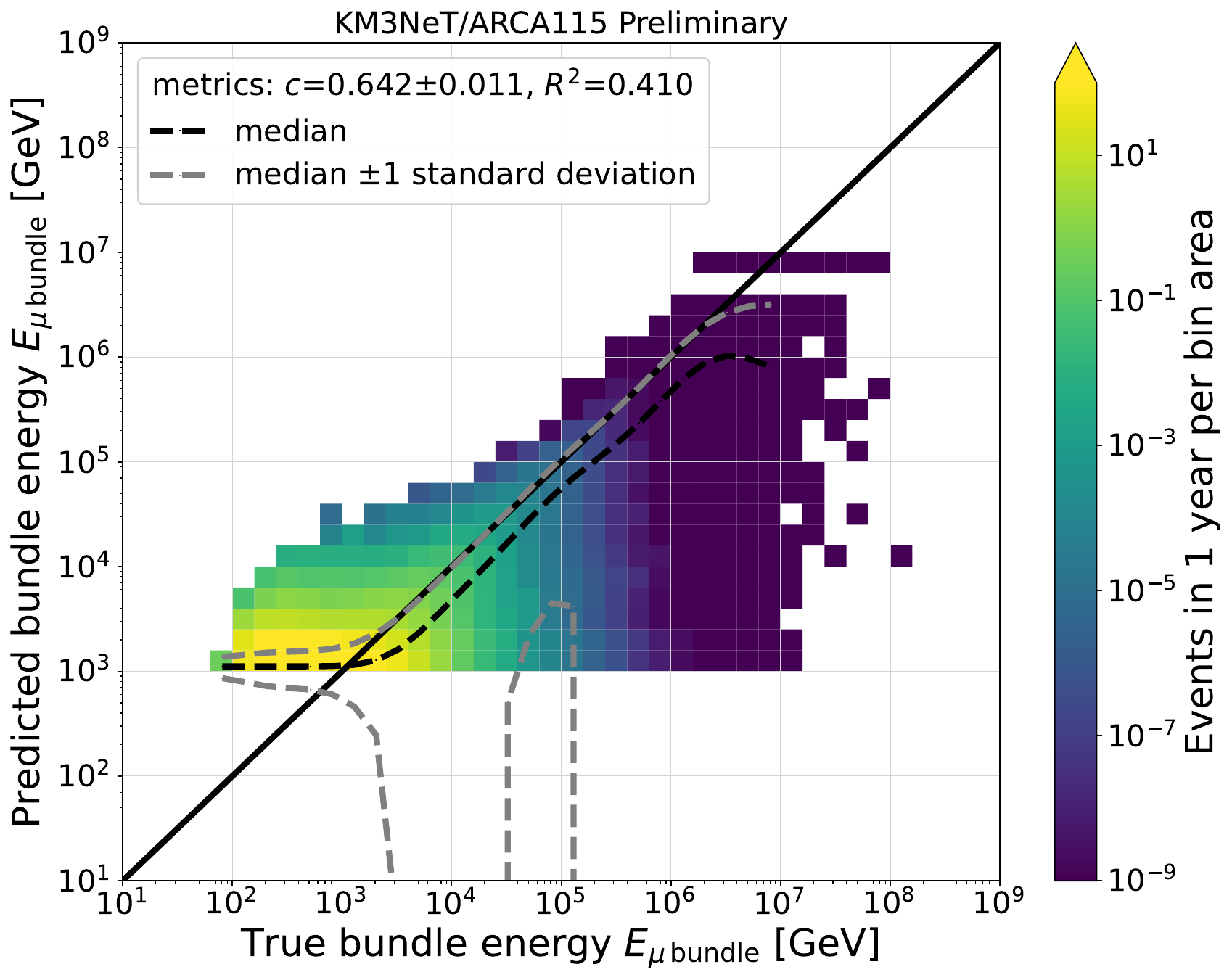}
         \caption{Only 3DMUON\_3DSHOWER\_trig\_hits.}
         \label{fig:only_most_important_feature}
     \end{subfigure}
     \hfill
     \begin{subfigure}[H]{0.49\textwidth}
         \centering
         \includegraphics[width=\textwidth]{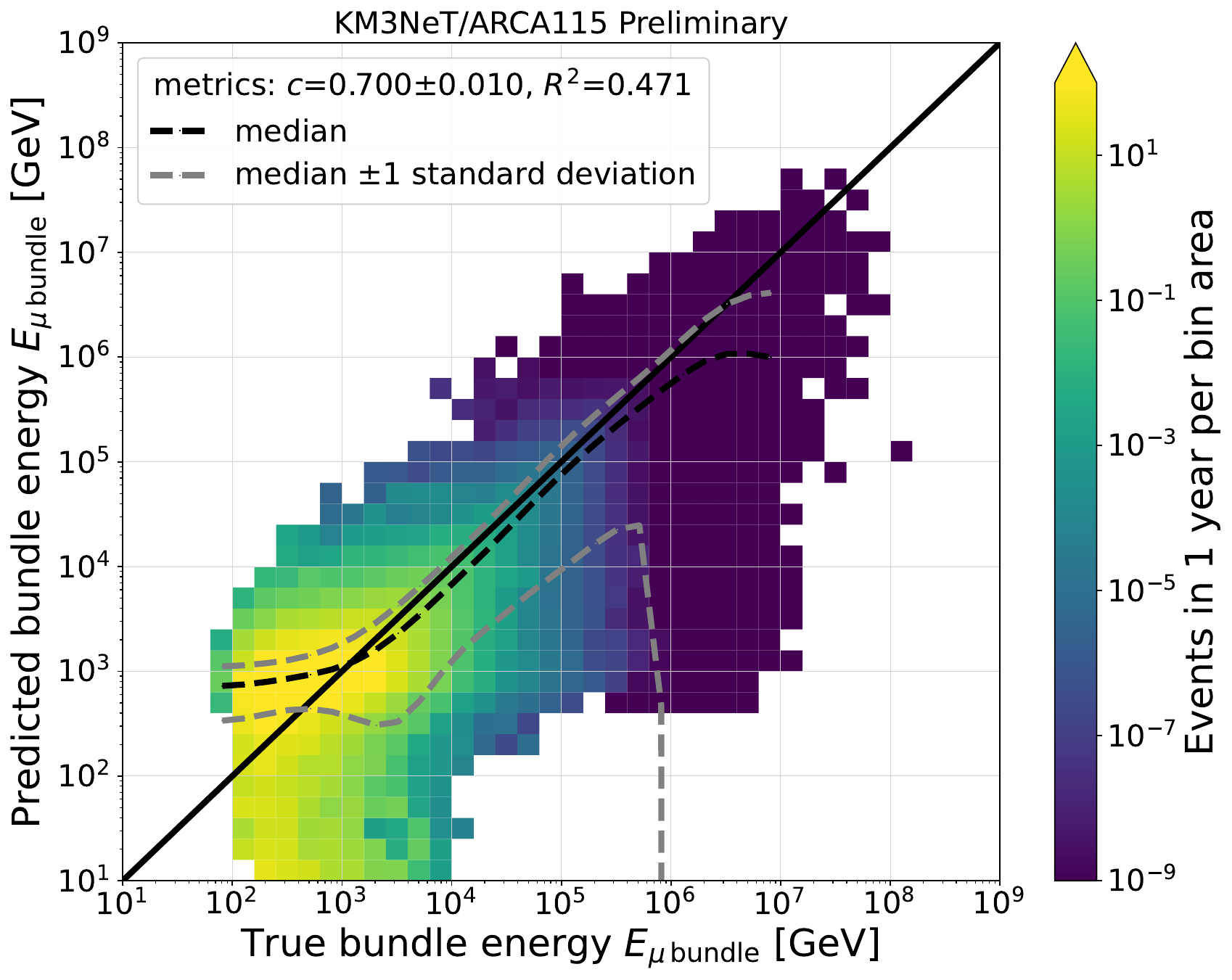}
         \caption{Selected features (positive importance).}
         \label{fig:selected_features}
     \end{subfigure}
    \caption{Comparison of predicted and true bundle energy for different feature selections for the ARCA115 configuration. The training target for the reconstruction was the bundle energy, not its logarithm. The logarithmic scale is only used for clearer visualisation.}
    \label{fig:feature_selection}
\end{figure}

The comparison between those four categories for ARCA115 is presented in Fig. \ref{fig:feature_selection}. The feature selection method requiring the positive importance of each feature provided the best results. Interestingly, even with a single feature (Fig. \ref{fig:only_most_important_feature}), it was possible to achieve a reasonable performance. There was no hope that this task could be reliably done just by counting the number of hits (the feature used is the number of triggered hits), however, it constituted an important sanity check, as hit counting is the basis of most of the energy reconstruction algorithms.

\begin{figure}[H]
    \centering
    \begin{subfigure}[H]{0.49\textwidth}
         \centering
         \includegraphics[width=\textwidth]{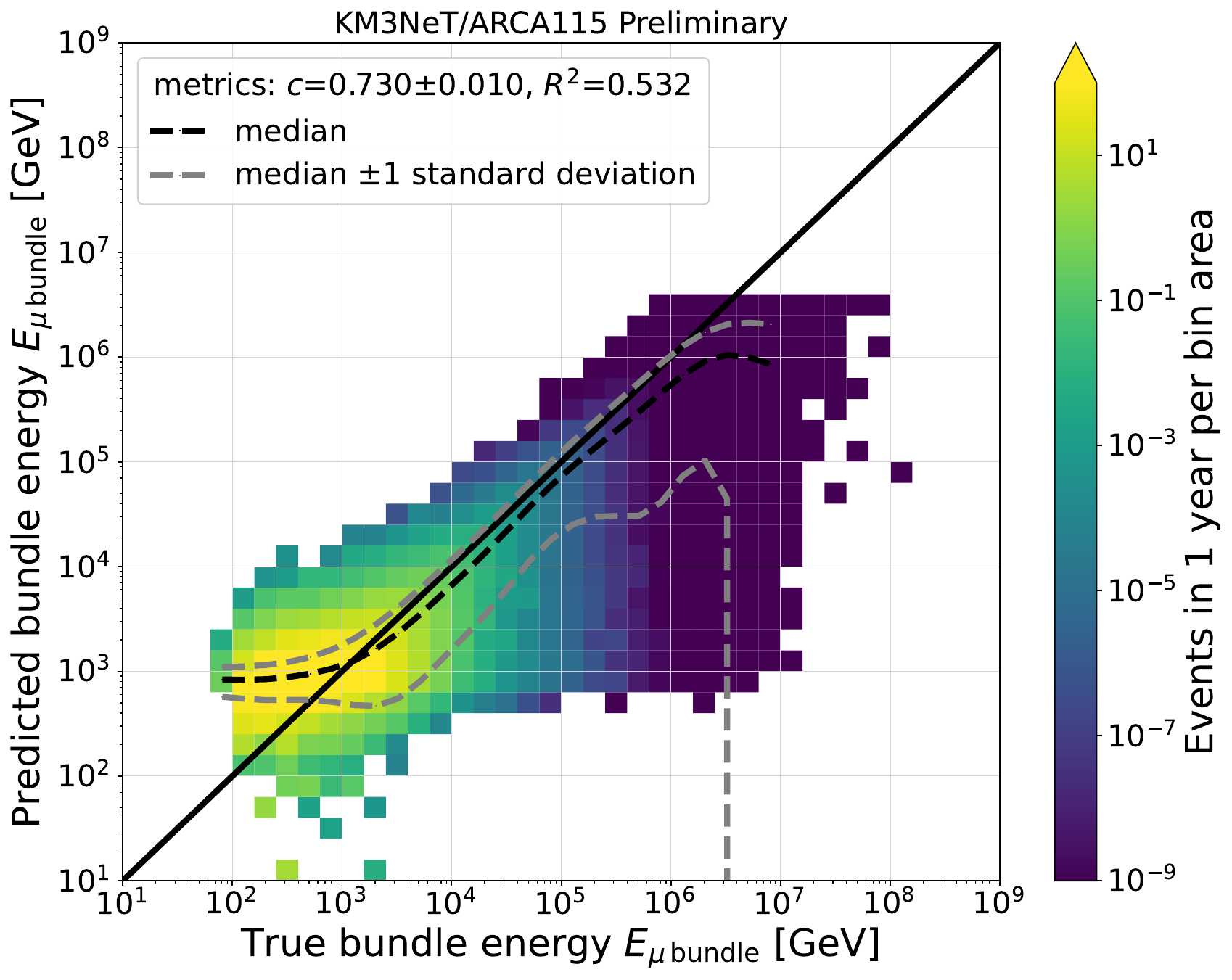}
         \caption{ARCA115.}
         \label{fig:E_bundle_A115}
     \end{subfigure}
     \hfill
     \begin{subfigure}[H]{0.49\textwidth}
         \centering
         \includegraphics[width=\textwidth]{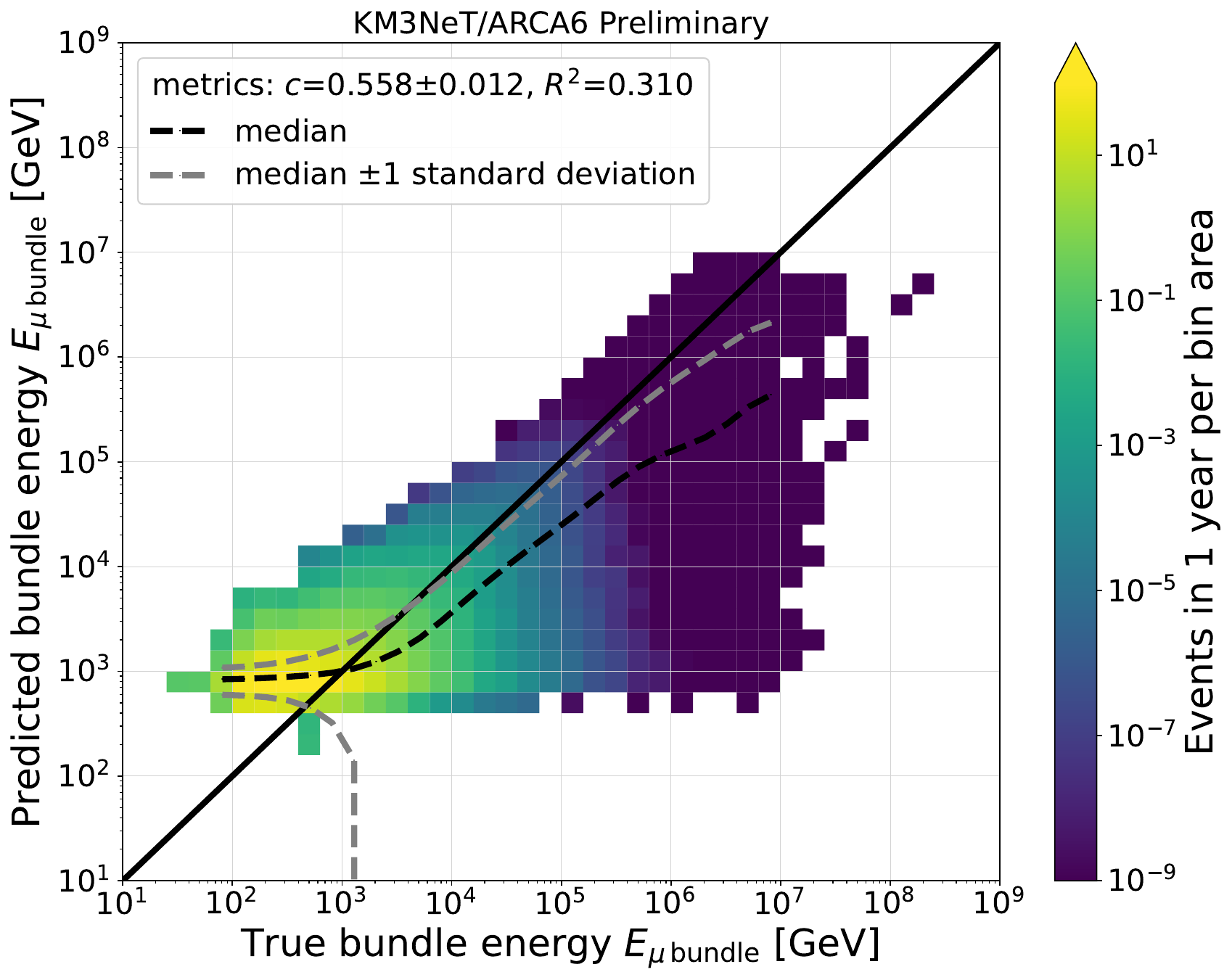}
         \caption{ARCA6.}
         \label{fig:E_bundle_A6}
     \end{subfigure}
     \newline
     \begin{subfigure}[H]{0.49\textwidth}
         \centering
         \includegraphics[width=\textwidth]{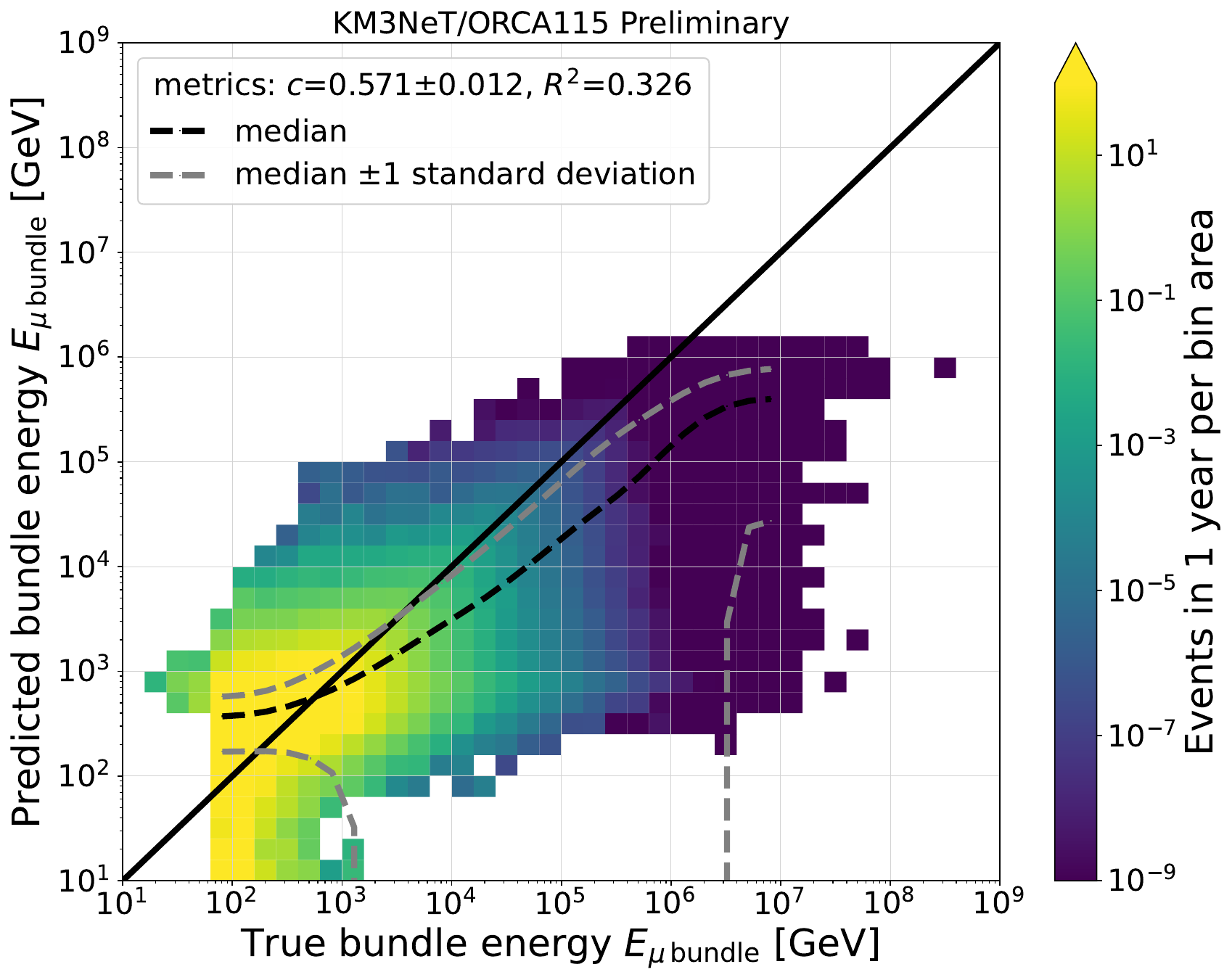}
         \caption{ORCA115.}
         \label{fig:E_bundle_O115}
     \end{subfigure}
     \hfill
     \begin{subfigure}[H]{0.49\textwidth}
         \centering
         \includegraphics[width=\textwidth]{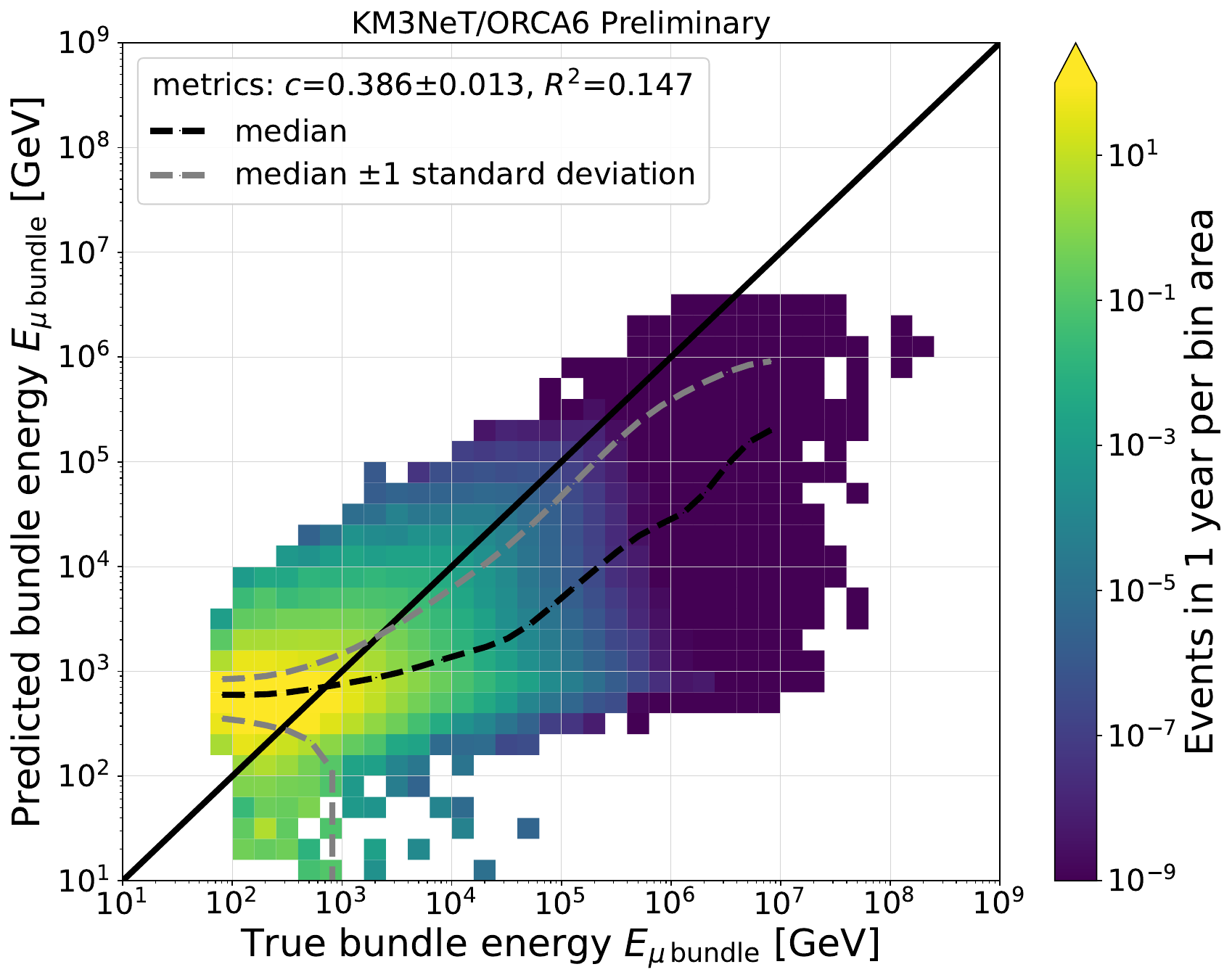}
         \caption{ORCA6.}
         \label{fig:E_bundle_O6}
     \end{subfigure}
    \caption{Comparison of reconstructed and true muon bundle energy for ARCA115 (a), ARCA6 (b), ORCA115 (c) and ORCA6 (d).}
    \label{fig:bundle_energy}
\end{figure}

After performing the hyperparameter tuning using the Optuna library \cite{Optuna}, the final results presented in Fig. \ref{fig:bundle_energy} were produced for the ARCA115, ARCA6, ORCA115 and ORCA6 configurations. Comparison of Fig. \ref{fig:E_bundle_A115} against Fig. \ref{fig:selected_features} shows that tuning is an important step, which can significantly improve the outcome. As expected, the bundle energy reconstruction is most accurate for ARCA115 --- this configuration instruments the largest volume and hence can observe the largest energy deposits. There is a threshold energy of about 1\,TeV below which the model cannot make reasonable predictions. Towards the highest energies, there is a growing underestimation of the true bundle energy. This might result from the smaller number of such events in the MC simulation and the smaller chance for such events to be fully contained within the can.

\begin{figure}[H]
    \centering
    \begin{subfigure}[H]{0.49\textwidth}
         \centering
         \includegraphics[width=\textwidth]{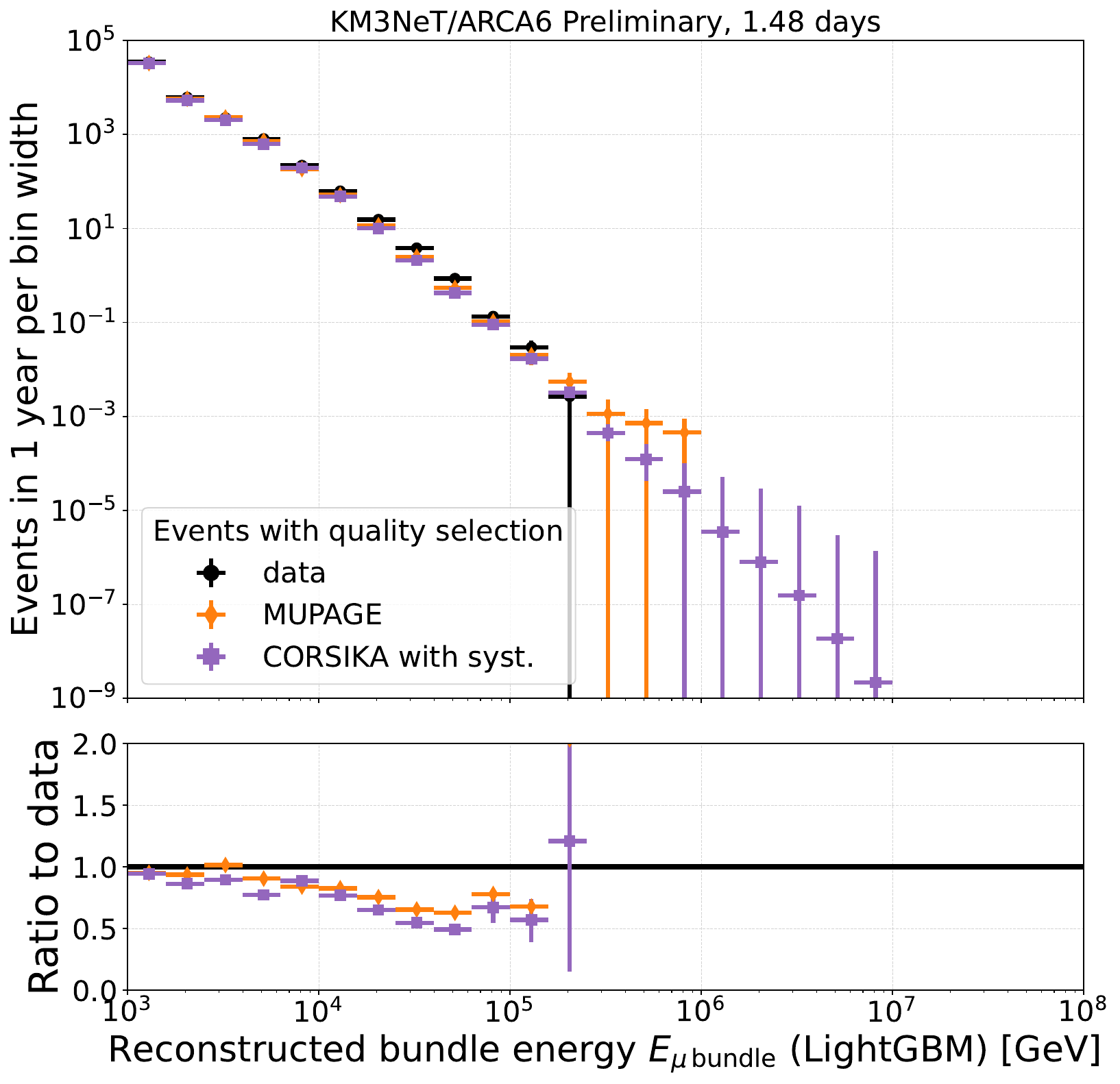}
         \caption{ARCA6.}
         \label{fig:data_vs_MC_Ebundle_A6}
     \end{subfigure}
     \hfill
     \begin{subfigure}[H]{0.49\textwidth}
         \centering
         \includegraphics[width=\textwidth]{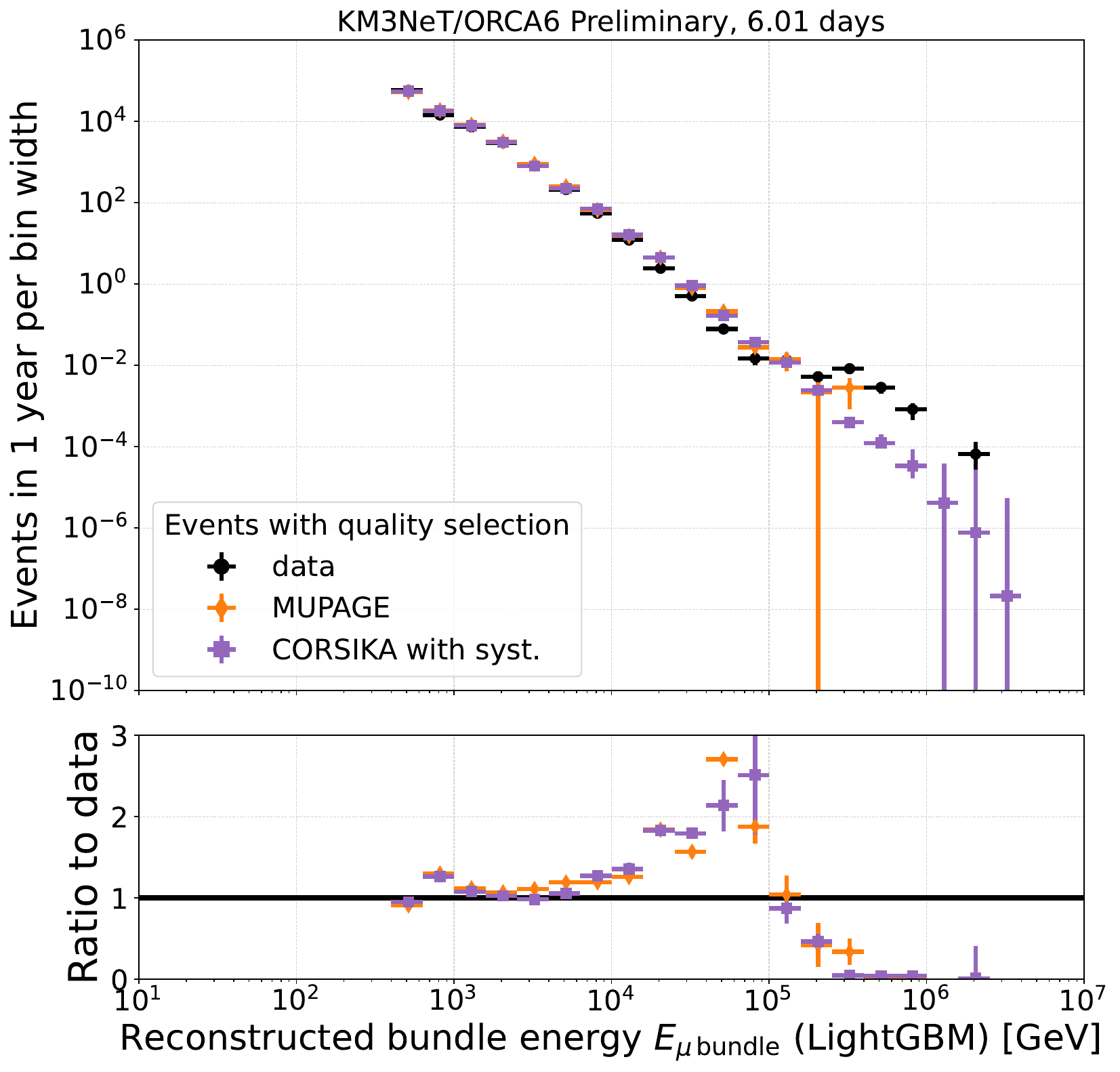}
         \caption{ORCA6.}
         \label{fig:data_vs_MC_Ebundle_O6}
     \end{subfigure}
    \caption{
    ML-based bundle energy reconstruction results for ARCA6 and ORCA6 data. The quality selection is described in detail in \cite{MyPhD}.
    }
    \label{fig:data_vs_MC_Ebundle}
\end{figure}

The comparison of the experimental data against the MC simulations for the reconstructed bundle energy is shown in Fig. \ref{fig:data_vs_MC_Ebundle}. The simulations reproduce the data within the few-TeV energy scale, however, at higher energies discrepancies are visible, similarly as in Fig. \ref{fig:data_vs_MC_multiplicity} and \ref{fig:data_vs_MC_Eprim}. The consistency between MUPAGE and CORSIKA behaviour suggests a common issue for both generators, whether it is the theoretical inputs, the way the simulated events are processed or something else.

\subsection{Primary cosmic ray energy}

\begin{figure}[H]
    \centering
    \begin{subfigure}[H]{0.49\textwidth}
         \centering
         \includegraphics[width=\textwidth]{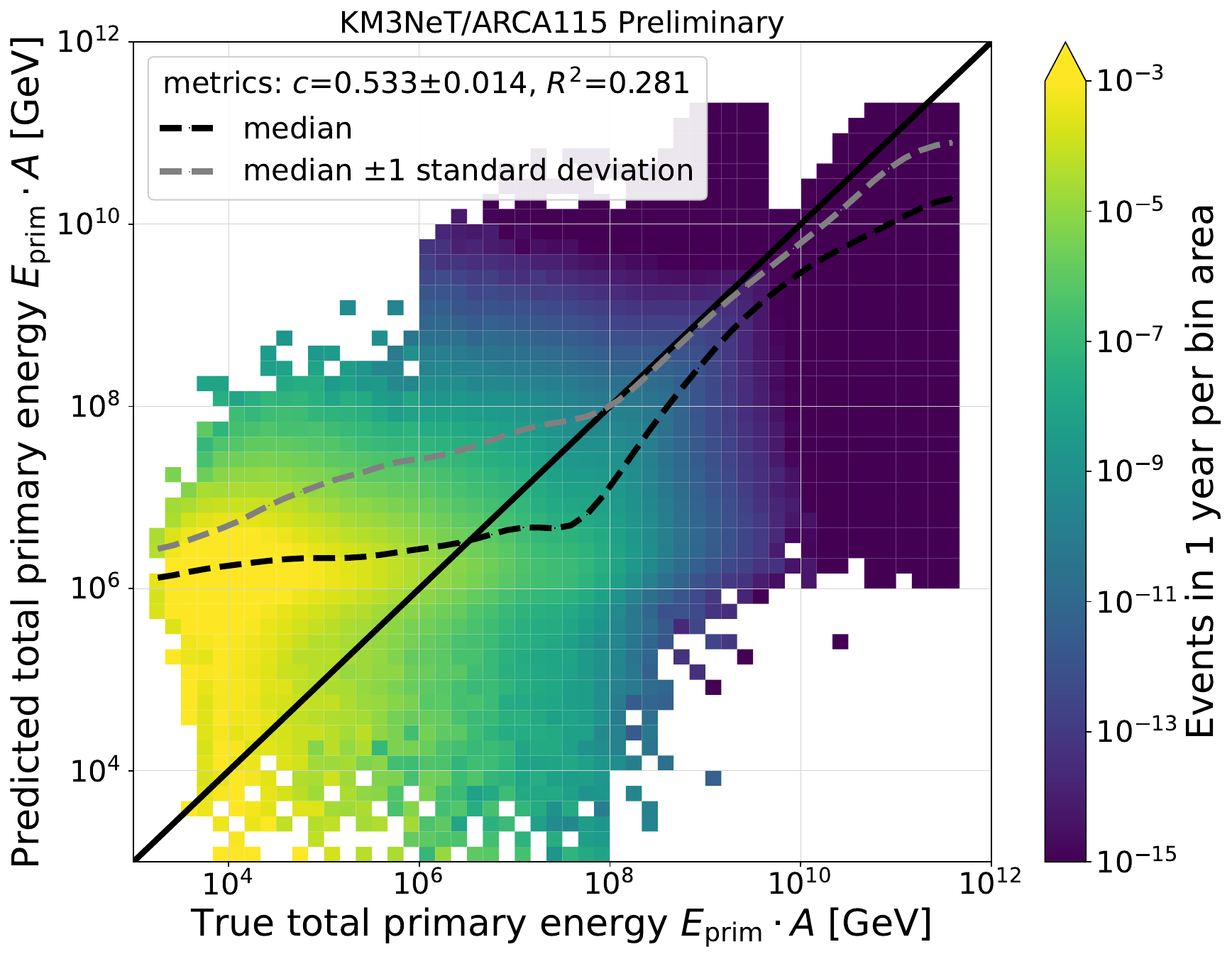}
         \caption{ARCA115.}
         \label{fig:E_prim_A115}
     \end{subfigure}
     \hfill
     \begin{subfigure}[H]{0.49\textwidth}
         \centering
         \includegraphics[width=\textwidth]{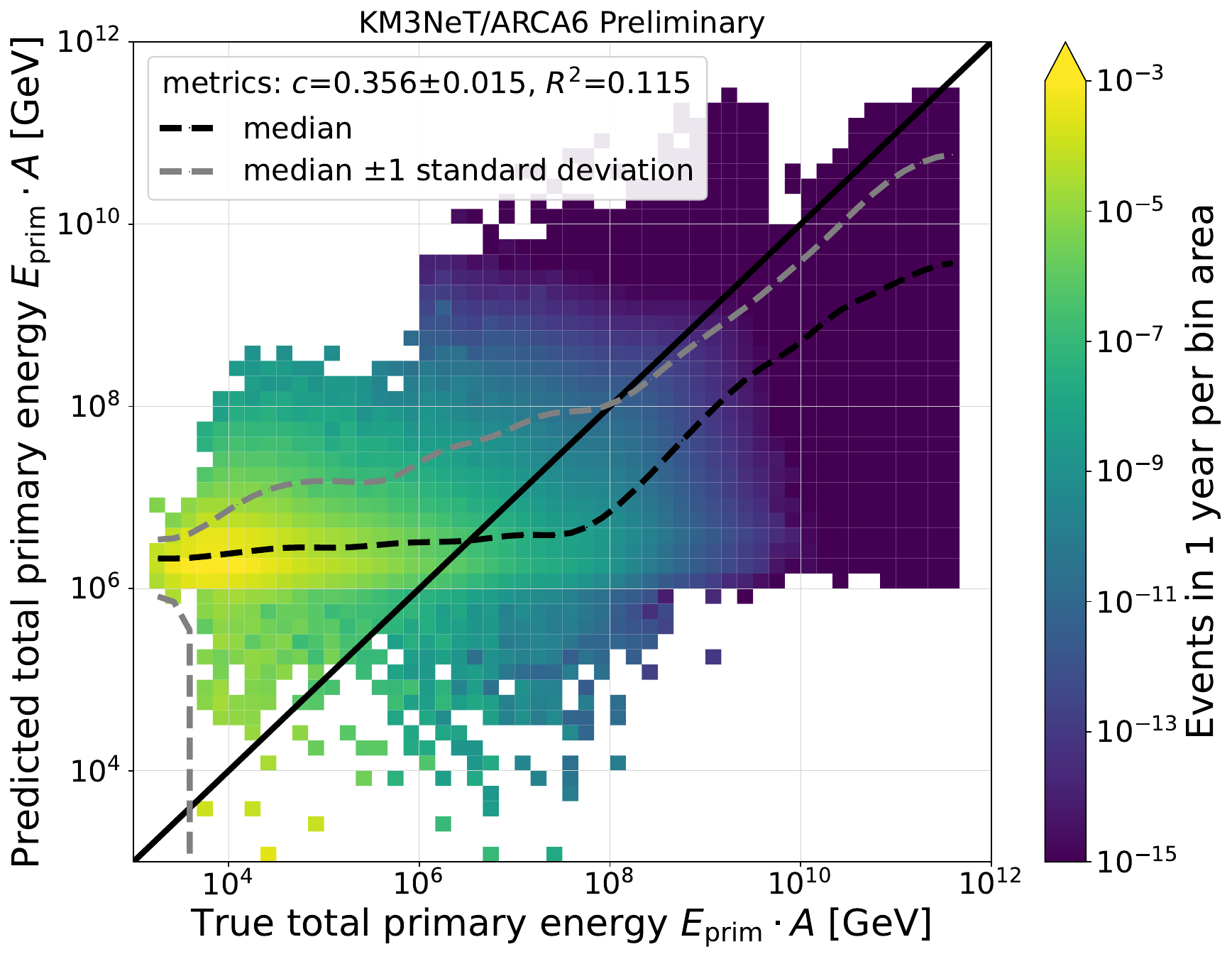}
         \caption{ARCA6.}
         \label{fig:E_prim_A6}
     \end{subfigure}
     \newline
     \begin{subfigure}[H]{0.49\textwidth}
         \centering
         \includegraphics[width=\textwidth]{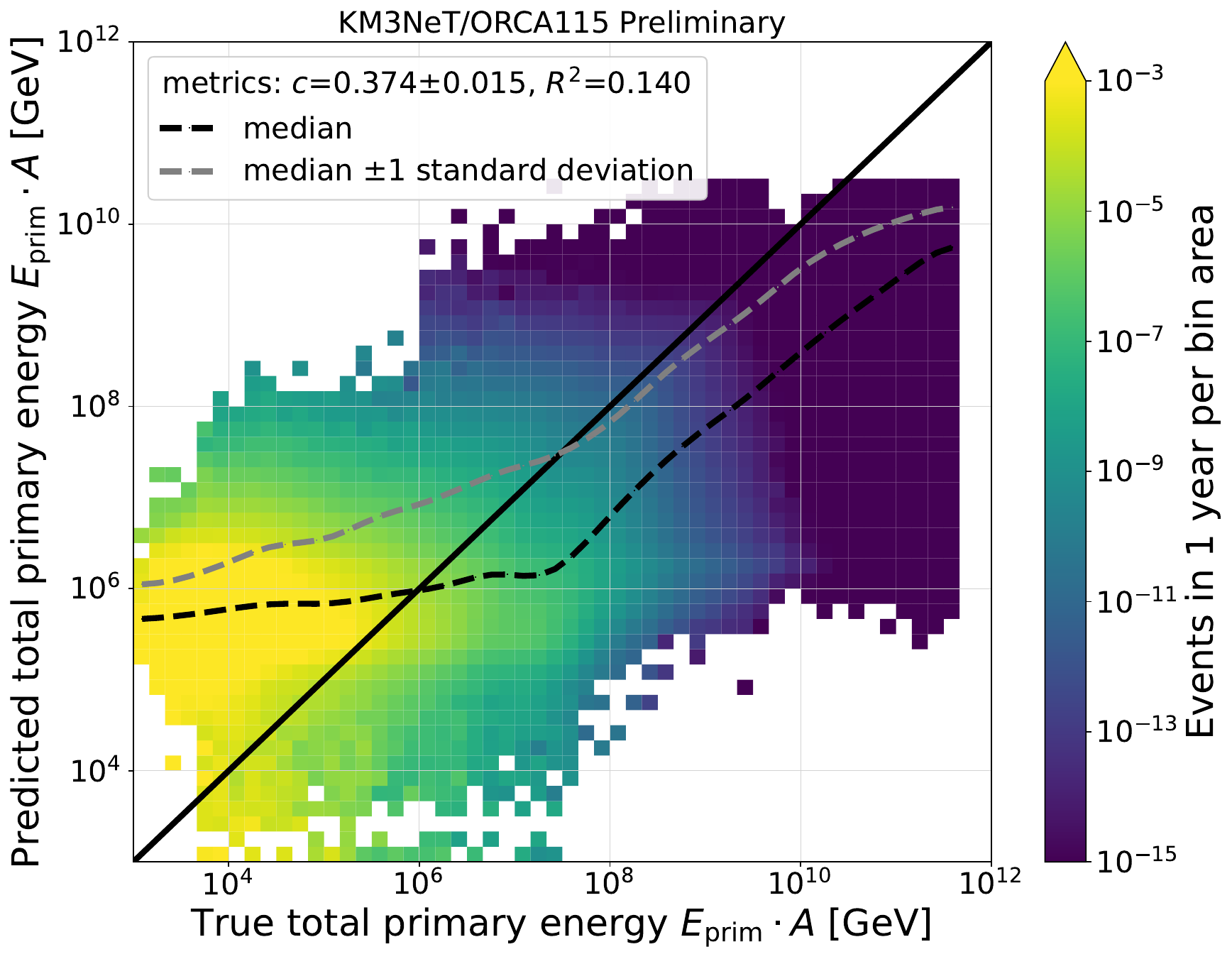}
         \caption{ORCA115.}
         \label{fig:E_prim_O115}
     \end{subfigure}
     \hfill
     \begin{subfigure}[H]{0.49\textwidth}
         \centering
         \includegraphics[width=\textwidth]{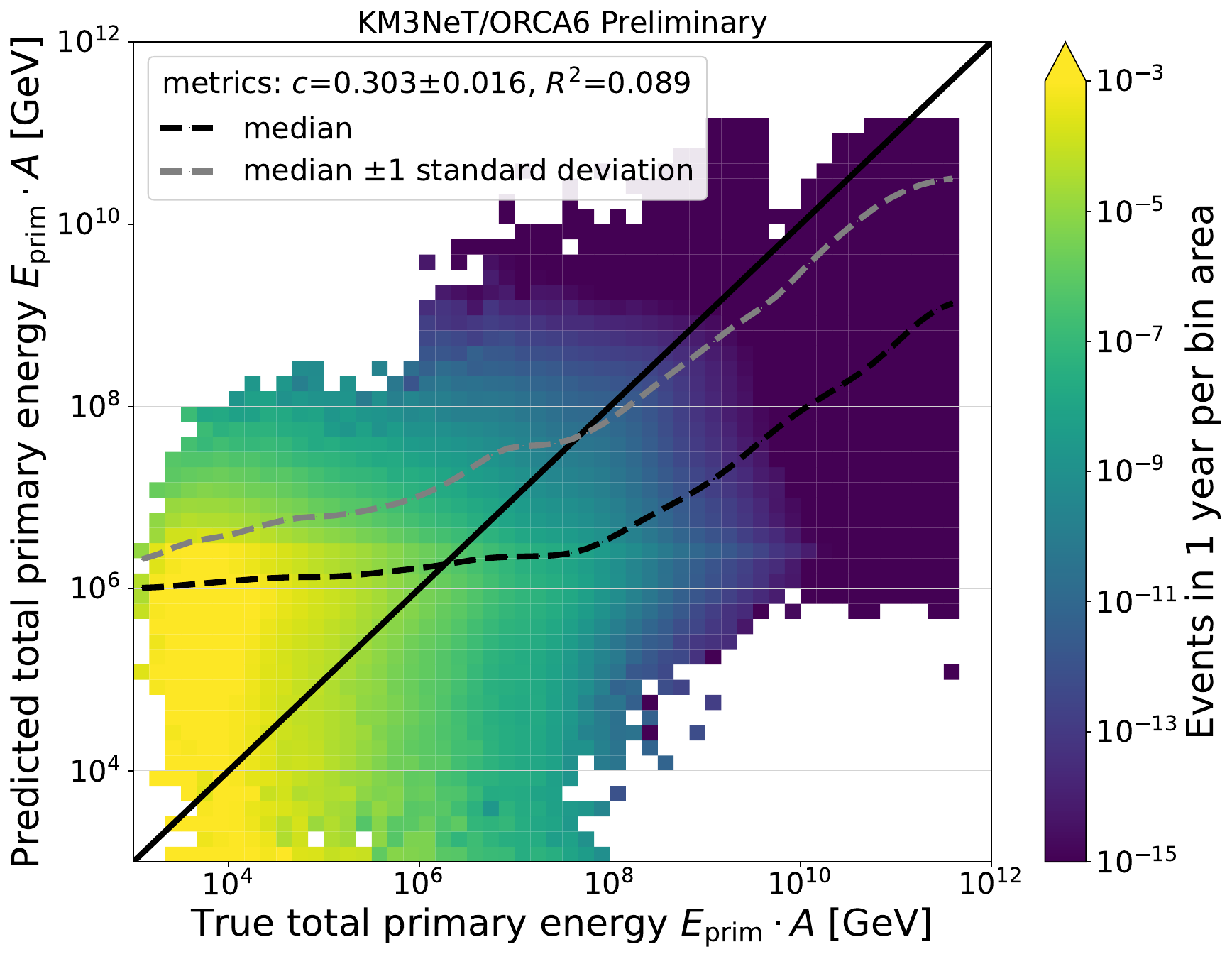}
         \caption{ORCA6.}
         \label{fig:E_prim_O6}
     \end{subfigure}
    \caption{Comparison of reconstructed and true total primary energy for ARCA115 (a), ARCA5(b) , ORCA115 and ORCA6 (d).}
    \label{fig:primary_energy}
\end{figure}

\begin{figure}[H]
    \centering
    \begin{subfigure}[H]{0.49\textwidth}
         \centering
         \includegraphics[width=\textwidth]{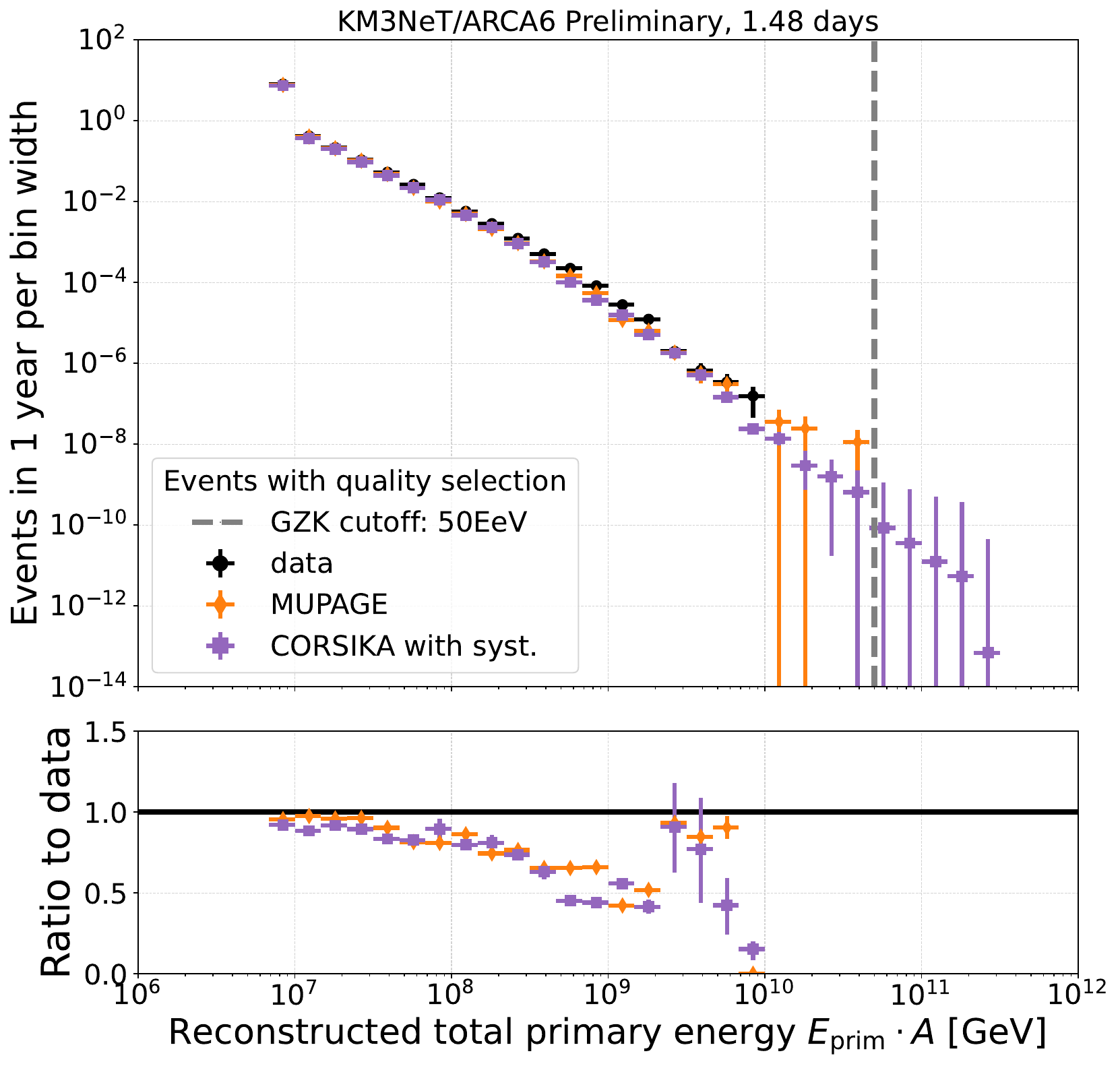}
         \caption{ARCA6.}
         \label{fig:data_vs_MC_Eprim_A6}
     \end{subfigure}
     \hfill
     \begin{subfigure}[H]{0.49\textwidth}
         \centering
         \includegraphics[width=\textwidth]{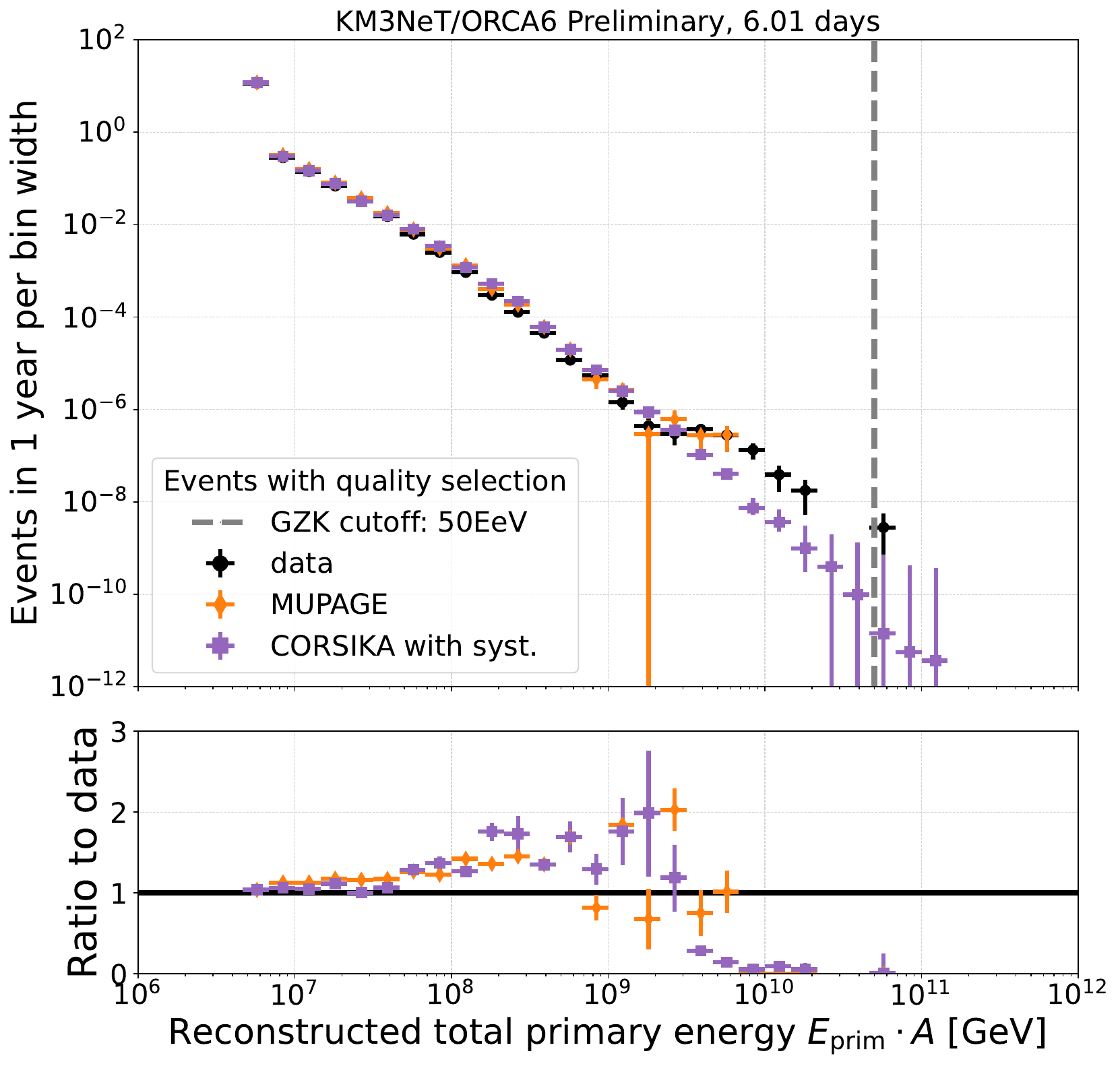}
         \caption{ORCA6.}
         \label{fig:data_vs_MC_Eprim_O6}
     \end{subfigure}
    \caption{
    ML-based primary CR energy reconstruction results for ARCA6 and ORCA6 data. The quality selection is described in detail in \cite{MyPhD}.
    }
    \label{fig:data_vs_MC_Eprim}
\end{figure}

The outcome of the previous subsection encouraged an attempt to reconstruct the primary CR energy. The same procedure was followed, with the same feature selection but a different hyperparameter tuning. Despite the bad performance, the results reported in Fig. \ref{fig:primary_energy} look promising above 100\,PeV, given how much of the original information is lost when the EAS has reached the detector. It suggests that KM3NeT may contribute to the primary CR measurements at the highest energies, notably to the study of the Greisen-Zatsepin-Kuzmin (GZK) cutoff \cite{GZK_1,GZK_2}. The data from ARCA6 and ORCA6 used in this work is insufficient to make any decisive statements about the (non-)observation of the GZK cutoff (see Fig. \ref{fig:data_vs_MC_Eprim}), both due to limited exposure time and detector size.

\subsection{Muon multiplicity}
\label{subsec:multiplicity}

\begin{figure}[H]
    \centering
    \begin{subfigure}[H]{0.49\textwidth}
         \centering
         \includegraphics[width=\textwidth]{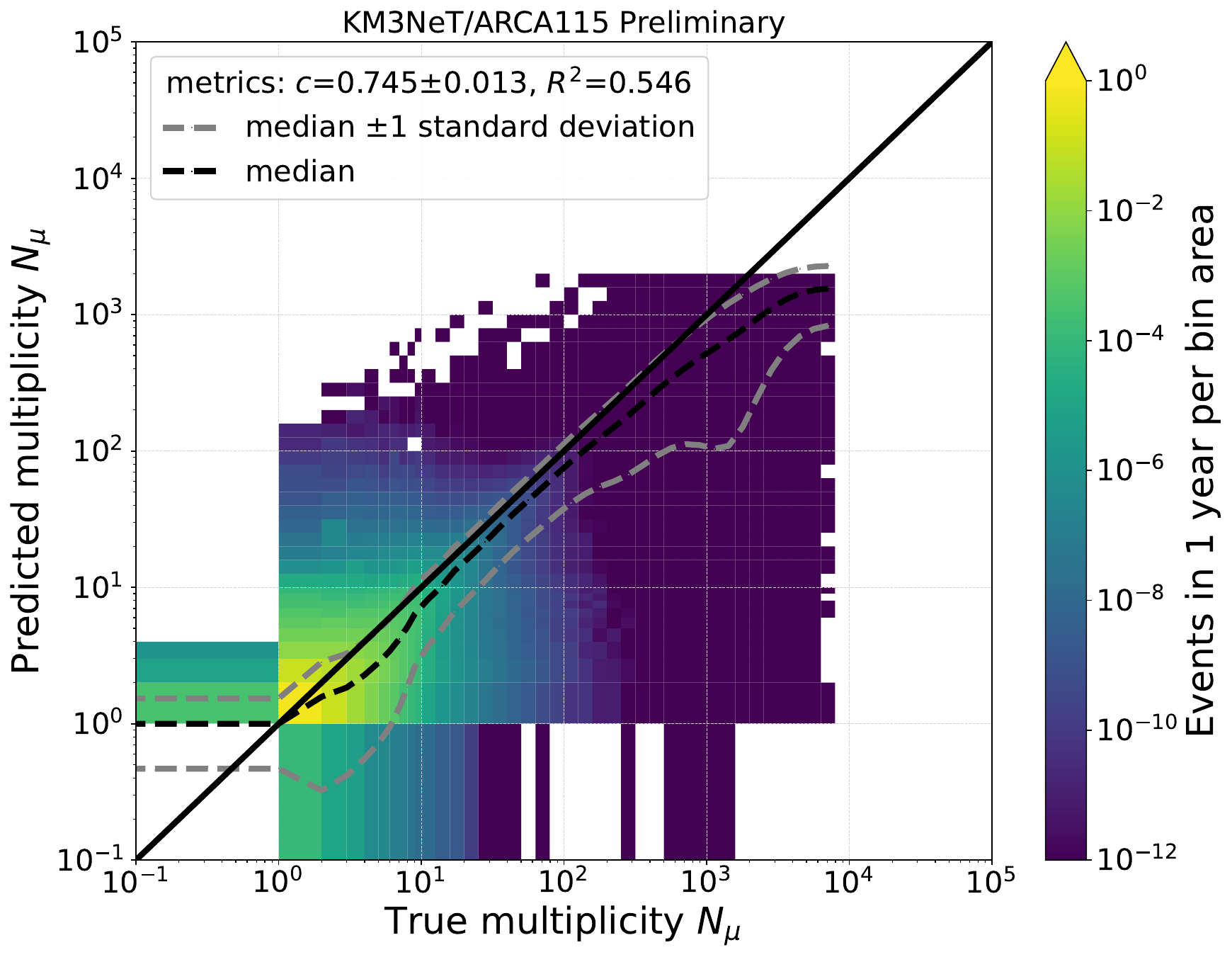}
         \caption{ARCA115.}
         \label{fig:N_mu_A115}
     \end{subfigure}
     \hfill
     \begin{subfigure}[H]{0.49\textwidth}
         \centering
         \includegraphics[width=\textwidth]{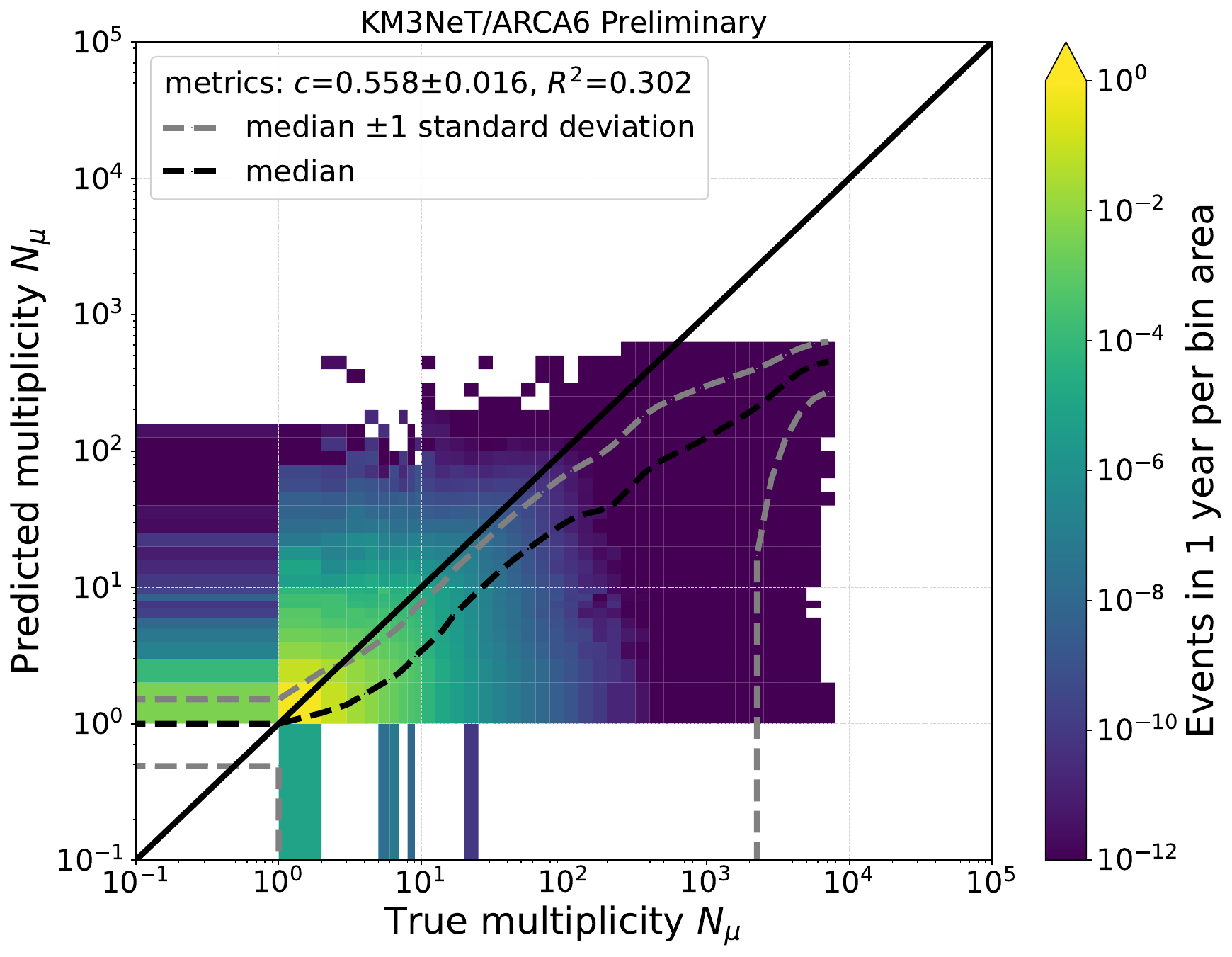}
         \caption{ARCA6.}
         \label{fig:N_mu_A6}
     \end{subfigure}
     \newline
     \begin{subfigure}[H]{0.49\textwidth}
         \centering
         \includegraphics[width=\textwidth]{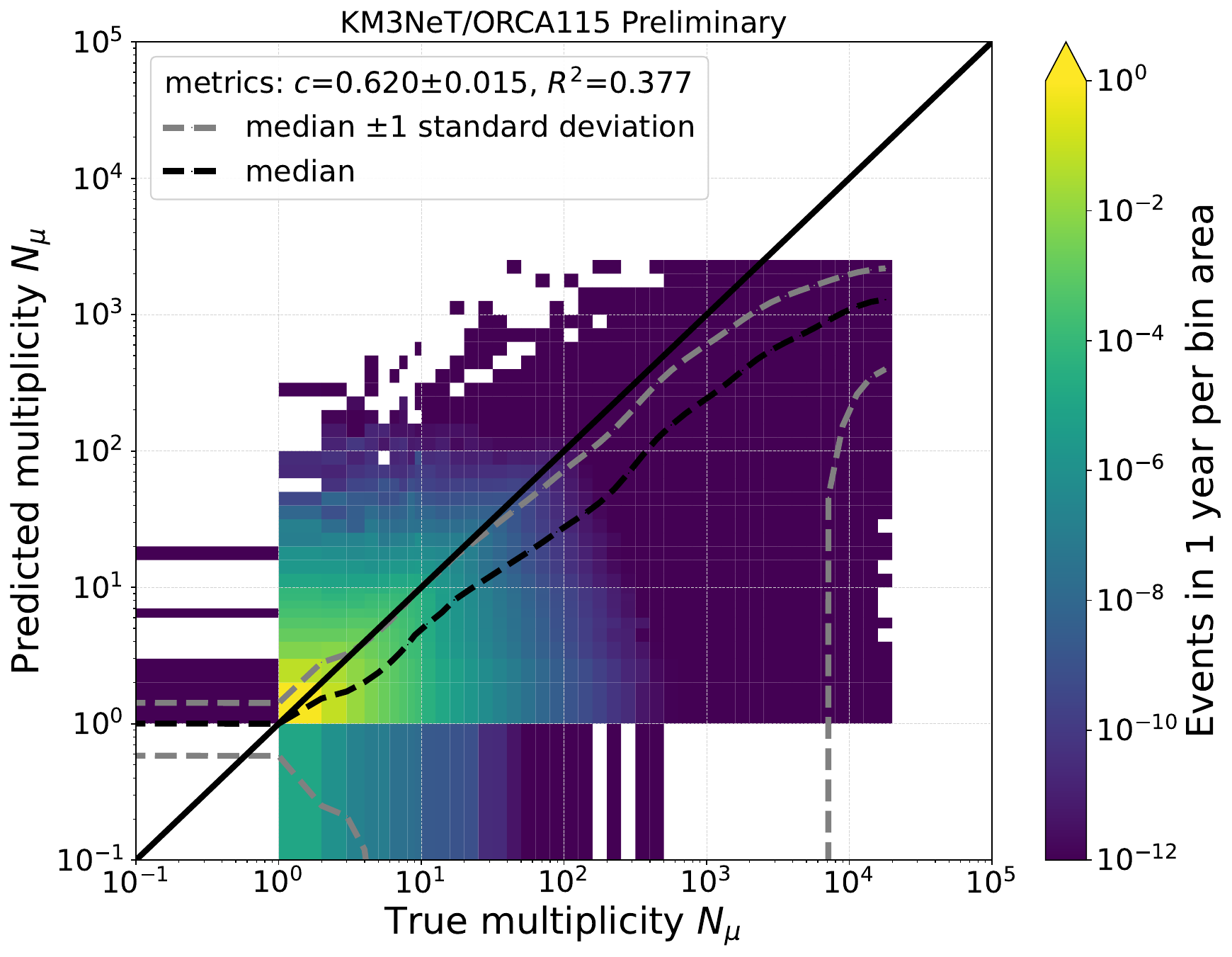}
         \caption{ORCA115.}
         \label{fig:N_mu_O115}
     \end{subfigure}
     \hfill
     \begin{subfigure}[H]{0.49\textwidth}
         \centering
         \includegraphics[width=\textwidth]{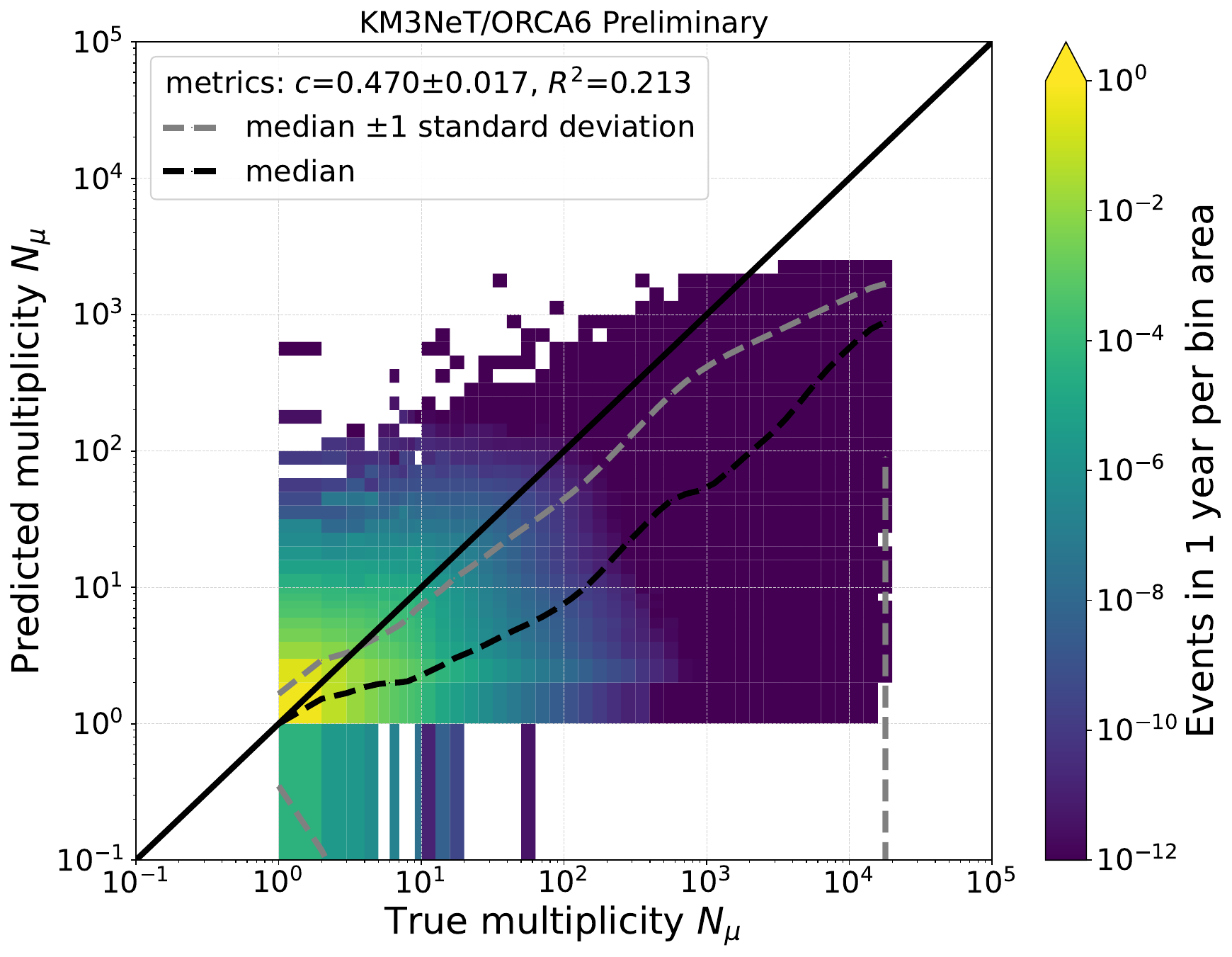}
         \caption{ORCA6.}
         \label{fig:N_mu_O6}
     \end{subfigure}
    \caption{
    Comparison of muon multiplicity reconstructed with the LigthGBM model against the true value for ARCA115 (a), ARCA5(b) , ORCA115 and ORCA6 (d). Both metrics were computed on the bin values, not individual events.
    }
    \label{fig:multiplicity}
\end{figure}

Counting the muons in the bundle is a challenging task. Some muons may not be energetic enough or only skim the edge of the sensitive volume, resulting in insufficient illumination of the PMTs. To account for this, the multiplicity definition has been adapted by counting only the muons with energy above 120\,GeV for ARCA and 1\,GeV for ORCA. An additional spatial criterion was found beneficial for ARCA6, requiring that the closest distance from the muon trajectory to the can centre is at most 269.4\,m and the muon pathlength inside the can is at least 240\,m. Details are provided in \cite{MyPhD}. Such a muon selection results in the possibility of having bundles with multiplicity equal to 0, which is indeed the case, except for ORCA6 (see Fig. \ref{fig:multiplicity}). Reconstructing such events proves troublesome for the same reasons that led to discarding the muons in the first place. Nevertheless, the overall performance is improved thanks to more accurate prediction for high-multiplicity events. The multiplicity reconstruction is more accurate for ARCA, thanks to the typical muon signature in a detector (elongated track). For each detector configuration, the best performance was observed for low (except for 0) and intermediate multiplicity, with a growing underprediction for events with a larger number of muons (with similar reasons as for the muon bundle energy).

\begin{figure}[H]
    \centering
    \begin{subfigure}[H]{0.49\textwidth}
         \centering
         \includegraphics[width=\textwidth]{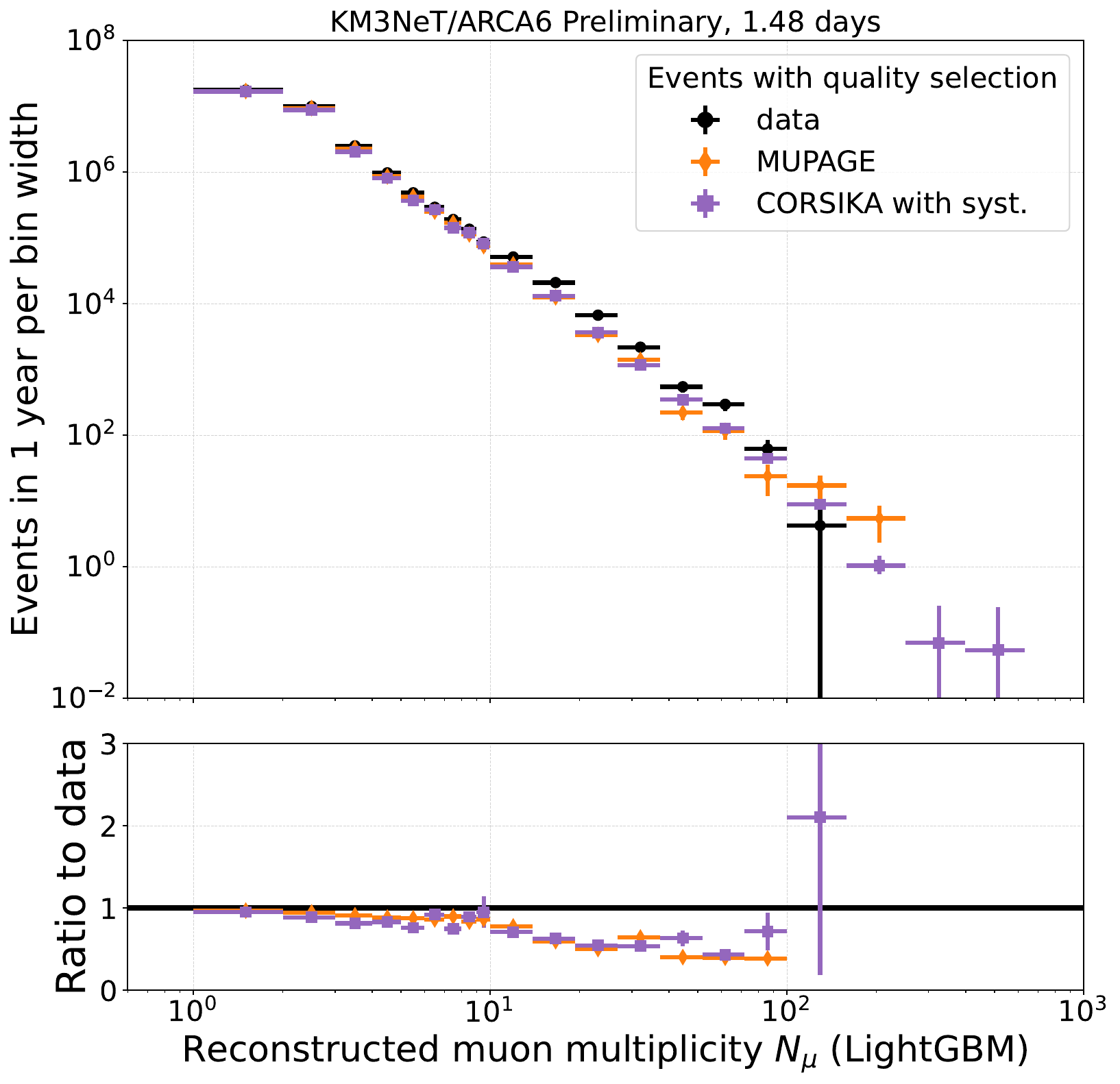}
         \caption{ARCA6.}
         \label{fig:data_vs_MC_N_mu_A6}
     \end{subfigure}
     \hfill
     \begin{subfigure}[H]{0.49\textwidth}
         \centering
         \includegraphics[width=\textwidth]{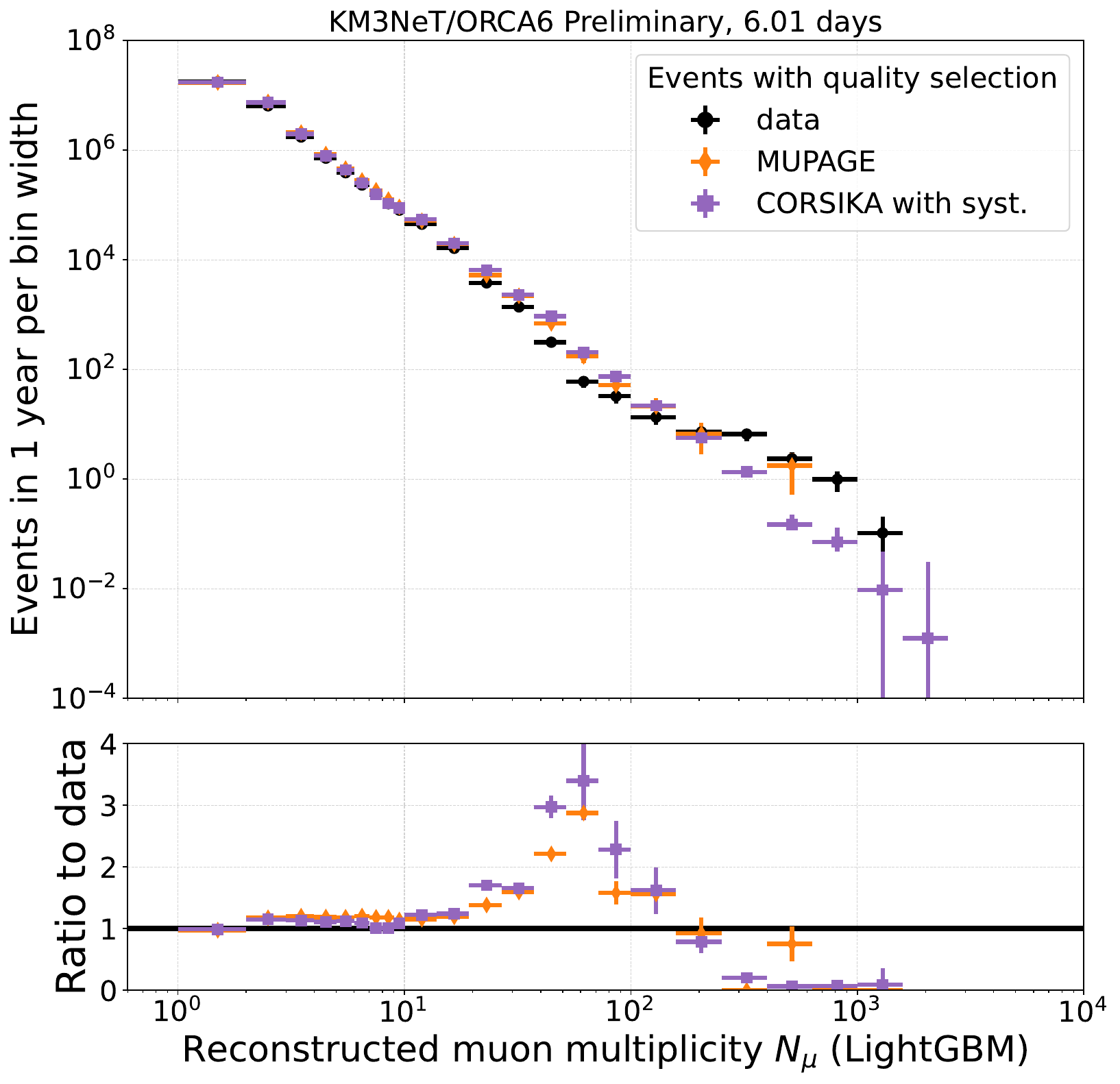}
         \caption{ORCA6.}
         \label{fig:data_vs_MC_N_mu_O6}
     \end{subfigure}
    \caption{
    ML-based energy reconstruction results for ARCA6 and ORCA6 data. The quality selection is described in detail in \cite{MyPhD}.
    }
    \label{fig:data_vs_MC_multiplicity}
\end{figure}

The application of the trained models to the experimental data is presented in Fig. \ref{fig:data_vs_MC_multiplicity}. There is good agreement between the data and MC simulations for bundles with multiplicity between 1 and 10. For MUPAGE and CORSIKA at higher multiplicities, the comparison between data and MC simulations indicates a need for further work on CR shower simulations to understand the underlying problem.

\section{Conclusion}

The partial KM3NeT detectors already collect data containing large numbers of muons created in air showers. They could be studied thanks to simulations generated with CORSIKA. Reconstruction of the energy of muon bundles and primary CR primaries using machine learning models is promising.

The potential for further improvement lies in achieving a better understanding of the discrepancies between the data and MC. The model training could potentially benefit from extracting features from even lower-level data to ensure minimal information loss. Considering other machine learning models, like graph neural networks \cite{Reck_GNN_ICRC}, might yield valuable results. For full detector configurations, an attempt to reconstruct individual muon tracks may be considered.

% \section{Acknowledgements} 
\begin{acknowledgements}
The authors acknowledge the financial support of the funding agencies:

%Belgium

Funds for Scientific Research (FRS-FNRS), Francqui foundation, BAEF foundation.

%Czech

Czech Science Foundation (GAČR 24-12702S);

%France

Agence Nationale de la Recherche (contract ANR-15-CE31-0020), Centre National de la Recherche Scientifique (CNRS), Commission Europ\'eenne (FEDER fund and Marie Curie Program), LabEx UnivEarthS (ANR-10-LABX-0023 and ANR-18-IDEX-0001), Paris \^Ile-de-France Region, Normandy Region (Alpha, Blue-waves and Neptune), France;

%Georgia

Shota Rustaveli National Science Foundation of Georgia (SRNSFG, FR-22-13708), Georgia;

%Germany (Max Planck Inst.) This work is part of the MuSES project which has received funding from the European Research Council (ERC) under the European Union’s Horizon 2020 Research and Innovation Programme (grant agreement No 101142396).

%Greece

The General Secretariat of Research and Innovation (GSRI), Greece;

%Italy

Istituto Nazionale di Fisica Nucleare (INFN) and Ministero dell’Universit{\`a} e della Ricerca (MUR), through PRIN 2022 program (Grant PANTHEON 2022E2J4RK, Next Generation EU) and PON R\&I program (Avviso n. 424 del 28 febbraio 2018, Progetto PACK-PIR01 00021), Italy; IDMAR project Po-Fesr Sicilian Region az. 1.5.1; A. De Benedittis, W. Idrissi Ibnsalih, M. Bendahman, A. Nayerhoda, G. Papalashvili, I. C. Rea, A. Simonelli have been supported by the Italian Ministero dell'Universit{\`a} e della Ricerca (MUR), Progetto CIR01 00021 (Avviso n. 2595 del 24 dicembre 2019); KM3NeT4RR MUR Project National Recovery and Resilience Plan (NRRP), Mission 4 Component 2 Investment 3.1, Funded by the European Union – NextGenerationEU,CUP I57G21000040001, Concession Decree MUR No. n. Prot. 123 del 21/06/2022;

%Morocco

Ministry of Higher Education, Scientific Research and Innovation, Morocco, and the Arab Fund for Economic and Social Development, Kuwait;

%The Netherlands

Nederlandse organisatie voor Wetenschappelijk Onderzoek (NWO), the Netherlands;

%Poland
The National Science Centre, Poland, grant number 2021/41/N/ST2/01177;
We gratefully acknowledge the funding support by program \enquote{Excellence initiative-research university} for the AGH University in Krakow as well as the ARTIQ project: UMO-2021/01/2/ST6/00004 and ARTIQ/0004/2021;
The grant \enquote{AstroCeNT: Particle Astrophysics Science and Technology Centre}, carried out within the International Research Agendas programme of the Foundation for Polish Science financed by the European Union under the European Regional Development Fund;
%Romania

Ministry of Research, Innovation and Digitalisation, Romania;

%Slovak Republic

Slovak Research and Development Agency under Contract No. APVV-22-0413; Ministry of Education, Research, Development and Youth of the Slovak Republic;

%Spain

MCIN for PID2021-124591NB-C41, -C42, -C43 and PDC2023-145913-I00 funded by MCIN/AEI/10.13039/501100011033 and by \enquote{ERDF A way of making Europe}, for ASFAE/2022/014 and ASFAE/2022 /023 with funding from the EU NextGenerationEU (PRTR-C17.I01) and Generalitat Valenciana, for Grant AST22\_6.2 with funding from Consejer\'{\i}a de Universidad, Investigaci\'on e Innovaci\'on and Gobierno de Espa\~na and European Union - NextGenerationEU, for CSIC-INFRA23013 and for CNS2023-144099, Generalitat Valenciana for CIDEGENT/2018/034, /2019/043, /2020/049, /2021/23, for CIDEIG/2023/20 and for GRISOLIAP/2021/192 and EU for MSC/101025085, Spain;

%UAE

Khalifa University internal grants (ESIG-2023-008 and RIG-2023-070), United Arab Emirates;

%UK

The European Union's Horizon 2020 Research and Innovation Programme (ChETEC-INFRA - Project no. 101008324).
\end{acknowledgements}
% \begin{acknowledgements}

%  The National Science Centre, Poland, grant number 2021/41/N/ST2/01177;
%  The grant "AstroCeNT: Particle Astrophysics Science and Technology Centre", carried out within the International Research Agendas programme of the Foundation for Polish Science financed by the European Union under the European Regional Development Fund;
%  We gratefully acknowledge the funding support by program "Excellence initiative-research university" for the AGH University in Krakow as well as the ARTIQ project: UMO-2021/01/2/ST6/00004 and ARTIQ/0004/2021.
% \end{acknowledgements}

% \nocite{*} %REMOVE
\bibliographystyle{cs-agh}
\bibliography{bibliography}
\end{document}